\documentclass[twocolumn, twoside, english, floatfix, a4paper, longbibliography, prb, superscriptaddress]{revtex4-2}
\usepackage[T1]{fontenc}
\usepackage[utf8]{inputenc}
\usepackage{graphicx}
\usepackage{float}
\usepackage{color}
\usepackage{amsbsy, amsmath, amssymb}
\usepackage{babel}

\begin{document}
\title{Interface modes in planar one-dimensional magnonic crystals}
\author{Szymon Mieszczak}
\email{szymon.mieszczak@amu.edu.pl}
\affiliation{Institute of Spintronics and Quantum Information, Faculty of Physics, Adam Mickiewicz University,
Pozna\'{n}, Poland}
\author{Jaros\l aw W. K\l os}
\email{klos@amu.edu.pl}
\affiliation{Institute of Spintronics and Quantum Information, Faculty of Physics, Adam Mickiewicz University,
Pozna\'{n}, Poland}

\begin{abstract}
We present the concept of Zak phase for spin waves in planar magnonic crystals and discuss the existence condition of interface modes localized on the boundary between two magnonic crystals with centrosymmetric unit cells. Using the symmetry criterion and analyzing the  logarithmic derivative of the Bloch function, we study the interface modes and demonstrate the bulk-to-edge correspondence. Our theoretical results are verified numerically and extended to the case in which one of the magnonic crystals has a non-centrosymmetric unit cells. We show that by shifting the unit cell, the interface modes can traverse between the band gap edges. Our work also investigate the role of the dipolar interaction, by comparison  the systems both with exchange interaction only and combined dipolar-exchange interactions.
\end{abstract}
\maketitle

\section{introduction}
Band structure is a distinctive feature of wave excitations in periodic structures. Solutions of the wave equations in a periodic medium, Bloch waves are characterized by the quasimomentum $\hbar \boldsymbol{k}$, related to the wavevector $\boldsymbol{k}$.
Adiabatic changes of the wavevector in the momentum space lead to the acquisition of a geometrical phase. Introduced by M. Berry\cite{berry_quantal_1984}, this phase is related to the topological invariants that distinguish the topological classes of the system. For Bloch waves $\Phi_k(x)$ propagating in a periodic 2D or 3D medium this role is played by Chern numbers\cite{Chern, KANE20133, Silveirinha15}, which are determined for successive bands from the Berry phases calculated along a closed loop in the momentum space. In a 1D system a loop for the Berry phase can be realized by sweeping the wavenumber $k$ across the  $1^{\rm st}$ Brillouin zone (i.e., in the range $[-\pi/a,\pi/a]$, where $a$ is the period of the structure). Then, we take advantage of the periodicity of the Bloch function in the reciprocal space: $\Phi_k(x)=\Phi_{k+2\pi/a}(x)$. Referred to as the Zak phase\cite{Zak82, Zak89}, this phase characterizes each band of a 1D crystal due to the lack of degeneration in 1D systems. The Zak phase can be altered by changing other parameters of the system (e.g., by tuning its structural and material parameters) significantly enough to disturb the band structure resulting in band gap closing and reopening.

The Zak phase has an ambiguity related to the selection of the unit cell\cite{atala_direct_2013}. However, for a centrosymmetric unit cell it only takes on two well-defined values, which are $0$ and $\pi$. These values classify the bands in two categories and distinguish the types of band gaps\cite{Zak85, Zak89, Rhim_2017}. 
The classification can be based on the symmetry of the Bloch functions on the edges of the bands/gaps\cite{Kohn59, Zak85}. 
This can help establish the criteria for the existence of edge or interface modes\cite{Zak85} in terminated periodic structures, where bulk characteristics (symmetry of the bands and their Zak phases) correlate with surface parameters determining the existence of edge modes in the band gaps. 

Zak phase and edge modes have been the subject of investigation in 1D continuous systems in the form of layered media or periodically corrugated waveguides.
Various kinds of systems have been studied, including photonic crystals\cite{Wang_2019, Xiao14}, microwave systems\cite{Nakata2020, chen_2014}, plasmonic crystals\cite{Wang18} and phononic crystals\cite{Li}. 
It is worthy of notice that the definition and interpretation of surface parameters  can vary with system. 
Examples include the rate of decay of electron waves outside the crystal (e.g., in vacuum\cite{Zak85}), surface impedance for electromagnetic waves\cite{Xiao14}, or pinning parameter for spin waves\cite{Rychly2015}.
The Zak phase is measurable quantity\cite{atala_direct_2013} and is a powerful tool to predict the existence and to describe properties of surface/edge modes.

The studies on spin waves in magnonic crystals\cite{kruglyak10, Krawczyk14, klos21} reported to date, mostly address lattice models based on the Heisenberg Hamiltonian\cite{Owerre_2017, Pershoguba18, Mook14} or the Landau-Lifshitz equation, but discretized in the second-quantization approach to the Bogliubov-de Gennes Hamiltonian\cite{Shindou2013}.
They strongly indicate the importance of the Dzyaloshinskii–Moriya interaction and dipolar interaction for the occurrence of non-zero Chern numbers. 
A general discussion of the topological origin of magnetostatic surface spin waves has been provided recently in Refs.~\cite{Mohseni19,Yamamoto, Liu20}.
Topological concepts can be used to reinterpret those studies of spin-wave defect and edge states in magnonic crystals\cite{kruglyak2006, Klos13, Lisenkov2014, Rychly17, Gallardo18, Osokin2018, Zhou22} which discuss the existence of localized states in terms of symmetry criteria\cite{Zak85}.

In this paper we demonstrate that 
(i) the same standard formula for the Zak phase as used for electronic states\cite{berry_quantal_1984}
applies to both exchange and exchange-dipolar spin waves in 1D planar magnonic crystals; 
(ii) in magnonic crystals with centrosymmetric unit cell the Zak phase can be determined by a symmetry-related criterion, and the values of the Zak phase for successive bands can be used to investigate the existence of interface states on the boundary between two semi-infinite magnonic crystals; 
(iii) the calculations done for the fictitious model, where the dipolar interactions were switched off, shows a close analogy to the electronic system;
(iv) the numerical calculations performed for a realistic, dipolar-exchange system confirm the theoretical predictions regarding the existence of interface spin-wave modes.
\section{Structure and model}
\subsection{Structure}
\begin{figure}[p]
\begin{centering}
\includegraphics[width=1\columnwidth]{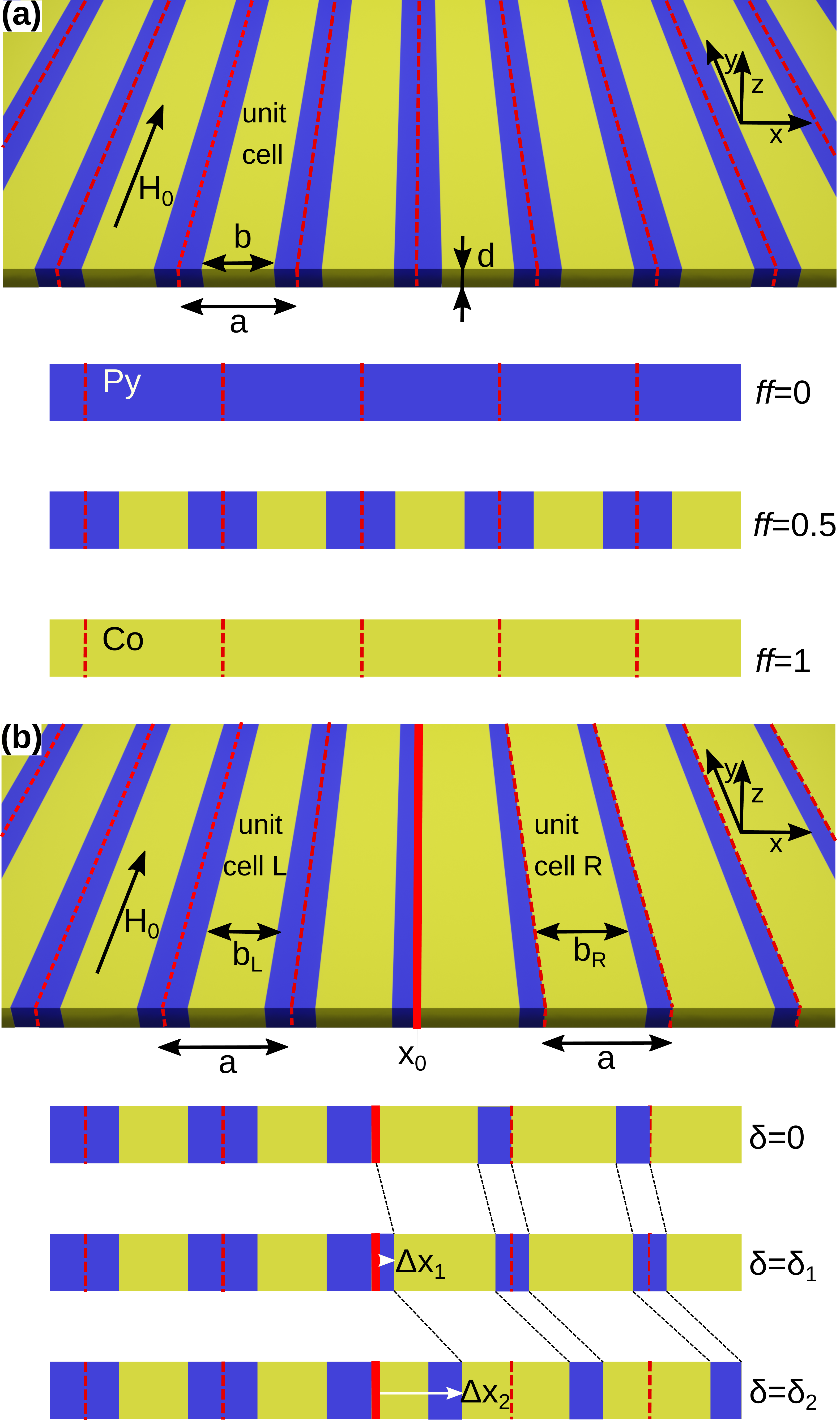}
\caption{(a) The geometry of the one-dimensional magnonic crystal. Red dashed lines mark the edges of centrosymmetric unit cells of the size $a$ (lattice constant). Yellow and blue colors distinguish the strips made of Co and Py of the width: $b$ and $a-b$, respectively. The strips are thin $d\ll a$, and the magnetic field $H_0$ is applied along the strips (static demagnetizing effects are absent). In our studies, we changed the width of both strips, keeping the lattice constant fixed (see the inset below (a)). By sweeping the bulk parameter (i.e., the filling fraction) $f\!f=b/a$ in the range $\left[0,1\right)$, we can tune the SW spectrum between the limits corresponding to the uniform layer of Py ($f\!f=0$) and Co ($f\!f=1$).
(b) Two semi-infinite magnonic crystals differing in filling fractions ($f\!f_L\ne f\!f_R$), interfaced at the edges of units cells (solid red line). For the magnonic crystal on the left (right) side, we chose a centrosymmetric (non-centrosymmetric) unit cell. The selection of the unit cell does not affect the spectrum of the infinite magnonic crystal but is important for the formation of interface states. The parameter $\delta=\Delta x/a=\left[0,1\right)$, describing the selection (i.e., the shift) of the unit cell, can then be treated as an interface parameter (see.the inset below (b)). The values $\delta_1$, $\delta_2$ (and $\Delta x_1$, $\Delta x_2$) correspond to two possible selections of centrosymmetric unit cell.
\label{fig:unitcell}}
\par\end{centering}
\end{figure}

We investigate the spin-waves (SWs) localized on the interface between two one-dimensional magnonic crystals (1D MCs). Each 1D MC is built from two strips, differing in magnetic parameters, that are arranged periodically in the plane, being in direct contact with each other. The structure of a single 1D MC is schematically presented in Fig.~\ref{fig:unitcell}(a). Such a system can be fabricated by lithographic techniques\cite{Wang09, Tacchi12, Choudhury16}, where two different materials can be deposited in separate areas, by the ion implantation, where magnetic anisotropy, magnetization saturation or exchange length can be changed in initially homogeneous magnetic layer\cite{Wawro18, Frackawiak20}, or by inducing a thermal gradient that suppress locally the magnetization saturation\cite{Chang18}. 
In our model, we consider two sets of parameters corresponding to widely used materials,
namely cobalt (Co) and permalloy (Py). 
The properties that are important for SW propagation are saturation magnetization $M_{\rm S}$ and exchange length $\lambda_{\rm ex}$. 
These parameters are equal to
$M_{\rm S,Co}=1445$~kA/m, $\lambda_{\rm ex,Co}=4.78$~nm, $M_{\rm S,Py}=860$~kA/m, $\lambda_{\rm ex,Co}=5.29$~nm. We assumed that both materials are amorphous, and there is no magnetocrystalline or surface magnetic anisotropy in our system. The strips are flat, i.e., their thickness $d$ is much smaller than their width. 
This assumption allows restricting our consideration to the SW fundamental mode, that is not quantized across the thickness. Additionally, we assume that our sample is saturated by an external magnetic field of the magnitude $\mu_{0}H_0=\mathrm{0.2}$~T oriented along the strips. For this magnetic configuration, the static demagnetizing field is equal to zero.

Two semi-infinite 1D MCs are jointed as it is presented in Fig.~\ref{fig:unitcell}(b). They are interfaced on the edges of their unit cells. The strips in both MCs are made of the same materials (Py and Co), and have the same period $a$ and thickness $d$. The structures on both sides of the interface differ only by: (i) the filling fractions $f\!f$ -- the ratio between the width of Co strip $b$ and the period $a$ ($f\!f=b/a$) and (ii) the selection of the unit cell  -- described by the parameter $\delta$. In the 1D crystal, the unit cell of the width $a$ can be shifted by arbitrary distance in the range $\Delta x=[0,a)$ ($\Delta x =0$ denotes unit cell with the whole Co(Py) strip on the left(right) side of unit cell). This selection does not affect the spectrum of infinite crystal (i.e., the band structure of propagating modes) but is important for the existence of surface/interface modes in the structures terminated at the edge of the unit cell. 
The parameter $\delta=\Delta x/a$ has two distinguished values
equal to: $\delta_1=1/2-f\!f/2, \delta_2=1-f\!f/2$. For these values unit cell becomes centrosymmetric. We calculated the Zak phase and logarithmic derivative of Bloch function for $\delta_1$, where both MCs have the same type of symmetry, i.e., Co strip in the middle of the unit cell.
Please also note that the cells for $\delta=0$ and $\delta=1$ are equivalent. 
In our studies, we are investigating the existence of SW modes localized on the interface between two 1D MC in the function of bulk parameter $f\!f$ and interface parameter $\delta$. 
One can imagine the system design based on modulation of some other bulk parameter, like value of external magnetic field\cite{Topp10, Mamica19}, thickness modulation\cite{Mruczkiewicz14, Bessonov15, Langer19}, or interface/surface parameters, like the modifications of the structures close to the interface/surface\cite{Rychly17, Klos13}.

\subsection{Model for magnetization dynamics}
We describe the SW modes in 1D MCs using the classical approach, based on the Landau-Lifshitz equation (LLE), which is an equation of motion for spatially dependent magnetization vector $\boldsymbol{M}(\boldsymbol{r},t)$ in the effective magnetic field $ \boldsymbol{H}_{\rm eff}(\boldsymbol{r},t)$:
\begin{equation}
\partial_t\boldsymbol{M}=-\mu_{0}|\gamma|\boldsymbol{M}\times\boldsymbol{H}_{\rm eff},\label{eq:LL}
\end{equation}
where $\mu_{0}=4\pi\times10^{-7}$H/m is the permeability of vacuum and $|\gamma|=194.6$ GHz/T gyromagnetic ratio. In our case $\boldsymbol{H}_{\rm eff}$ is composed of the following terms:
\begin{equation}
\boldsymbol{H}_{\rm eff}(\boldsymbol{r},t)=\boldsymbol{H}_{0}(\boldsymbol{r})+\boldsymbol{H}_{\rm dm}(\boldsymbol{r},t)+\boldsymbol{H}_{\rm ex}(\boldsymbol{r},t).
\end{equation}
The symbols: $\boldsymbol{H}_{0}$,  $\boldsymbol{H}_{\rm dm}\left(\boldsymbol{r},t\right)$ and $\boldsymbol{H}_{\rm ex}\left(\boldsymbol{r},t\right)$ stand for external field, demagnetizing field and exchange field, respectively. The last two terms are both spatially and temporally dependent since they are related to dynamic exchange and dynamic dipolar interaction. 
The SWs are calculated in linear approximation, where the magnetization dynamics can be considered as precession motion around the static magnetic configuration $\boldsymbol{M}(\boldsymbol{r})\approx M_{\rm S}\hat{\boldsymbol{y}}$ with dynamic component $\boldsymbol{m}(\boldsymbol{r},t)=\boldsymbol{m}(\boldsymbol{r})e^{i\omega t}$ circulating harmonically in time, with the frequency $\omega$: $\boldsymbol{M}(\boldsymbol{r},t)=\boldsymbol{M}\left(\boldsymbol{r}\right)+\boldsymbol{m}(\boldsymbol{r})e^{i\omega t}$. We consider only the case where SWs propagate along the direction of periodicity $\hat{\boldsymbol{x}}$. Therefore, the SW amplitude $\boldsymbol{m}(x)=m_\parallel(x)\hat{\boldsymbol{x}} +m_\perp(x)\hat{\boldsymbol{z}}$ depends only on $x-$coordinate. In linear approximation the LLE (\ref{eq:LL}) has a form of the set of two ordinary linear differential equations with  periodic coefficients (see Supplementary Information, section 1). Therefore, according to the Floquet's theorem\cite{Teschl}, their solutions can be presented as Bloch function  $\boldsymbol{m}_k(x)=\boldsymbol{u}_k(x) e^{i k x}$ with two complex components $m_{k,\parallel}$, $m_{k,\perp}$ related to in-plane and out-of plane magnetization dynamics, respectively. The symbol $k$ stands for the wavenumber and $\boldsymbol{u}_k(x)=u_{k,\parallel}(x)\hat{\boldsymbol{x}} +u_{k,\perp}(x)\hat{\boldsymbol{z}}$ is periodic component of the Bloch function: $\boldsymbol{u}_k(x)=\boldsymbol{u}_k(x+a)$.

In this study, we consider SW spectra for two kind of effective field: (i) dipolar interaction are neglected; (ii) dipolar interactions are included. In the first case, we assumed that the unit cell has a size equal to $a=100$ nm, while in the second, $a=1000$ nm. In both cases, the thickness $d=20$ nm $\ll a$ and in the model, we assume an infinite length of strips that gives us an effectively 1D system.

\subsection{Interface states: bulk to edge correspondence}
We followed the work \cite{Zak89} to determinate the Zak phase as a topological characteristic of every ($n^{\rm th}$)  band of the dispersion relation:
\begin{equation}
    \theta_n=\Im \int_{-\pi/a}^{\pi/a}\frac{\int_{-a/2}^{a/2}\boldsymbol{u}_{n,k}^{*}\cdot\partial_{k}\boldsymbol{u}_{n,k}dx}{\int_{-a/2}^{a/2}\boldsymbol{u}_{n,k}^{*}\cdot\boldsymbol{u}_{n,k}dx}dk,\label{eq:Zak}
\end{equation}
for two MCs which were then joined at the common interface. 
In Appendix \ref{sec:App.A} we present a detailed discussion of the applicability of the formula (\ref{eq:Zak}) to SW. 
The value of the Zak phase depends on the selection of the unit cell\cite{atala_direct_2013}. For the centrosymmetric unit cell, the Zak phase takes two quantized values, either $0$ or $\pi$ and can be deduced from symmetry criterion of modes (see Appendix \ref{sec:App.B} for details). 

The necessary condition to observe the SW modes localized on the interface of two MCs (Fig.~\ref{fig:unitcell}(b)) is an overlapping some frequency gaps in the spectra of both MCs. This fact ensures the exponential decaying of the mode on both sides of the interface, with the rate $\pm k_i$. For centrosymmetric unit cell, the logarithmic derivative is real and has a constant sign within the gap\cite{Zak85} (see Appendix \ref{sec:App.B}). Therefore, the matching of the signs of logarithmic derivatives of Bloch function:
\begin{equation}
\rho \left( k\right) = \left. \partial_{x} \ln\left(m_{k,\alpha}\left(x\right)\right)\right|_{x=x_{0^\pm}}=\left. \frac{\partial_{x} m_{k,\alpha}(x)}{m_{k,\alpha}(x)}\right|_{x=x_{0^\pm}},\label{eq:rho}
\end{equation}
on both sides of the interface  between two MCs ($x=x_{0^\pm}$) is equivalent to the fulfillment the boundary conditions for $m_{k,\alpha}$  (for each component ($\alpha=\{\perp,\parallel\}$) .
These conditions allow finding the SW interface modes. 
It is worth noting that that we can limit our consideration only to one complex component of the Bloch function because $m_{k,\perp}/m_{k,\parallel} = C e^{-i \pi/2}$, where $C$ is real and has a constant sign, determined by the direction of precession ($C=1$ for purely exchange waves).

The relation between the sign of logarithmic derivative $\rho$ in the gap above $n^{\rm th}$ band and Zak phases $\theta_m$ can be written for MC of centrosymmetric unit cells as\cite{Xiao14} (see Appendix \ref{sec:App.B}):
\begin{equation}
   {\rm sgn}(\rho)=\pm (-1)^{n-1}\exp\sum_{m=1}^n i\theta_m,\label{eq:rho_theta}
\end{equation}
where $m=1,\ldots,n$ indexes all bands below the gap. 
The signs '+' and '-' in the formula (\ref{eq:rho_theta}) refers to two possible selection of the complex wavevector in the gap, which describes the mode decaying to the right ($k = k_r+i k_i$) and left ($k = k_r - i k_i$) in the crystal, respectively\cite{Kohn59} ($k_i>0$). The Eq. (\ref{eq:rho_theta}) relates the topological parameter (Zak phases) characterizing the bands of MC(s) with the boundary condition at the interface (expressed by the logarithmic derivative of the SW amplitude). 
The modes localized on the interface between two MCs (can) cannot exist when the signs of logarithmic derivative are the same (different) on both side of the interface (see the plots of $\rho$ for both MC in Appendix C). It means that the expression $(-1)^{n-1}\exp\sum_{m=1}^n i\theta_n$ must have opposite signs on both side of the interface to compensate the change of the sign related to different direction of decaying of interface modes for $x\rightarrow\pm\infty$  (i.e., the signs '$\pm$' at the beginning of the formula for  ${\rm sgn}(\rho)$).

The logarithmic derivative taken at the symmetry points of centrosymmetric unit cells has zeros and poles only at the edges of frequency gaps and $\rho$ is real inside the gaps (see Appendix B for details). It means that $\rho$ cannot change its sign inside the gap.
Therefore, the agreement of the signs of $\rho$ in common frequency gaps, and eventually the existence of interface modes, depends on qualitative features of both MCs spectra. More precisely, depends on the sequences of zeros and poles of $\rho$ at gaps/bands boundaries and its signs in successive gaps. We show in the Appendix B that mentioned qualitative changes in the the spectrum can be expresses as a $0\leftrightarrow\pi$ flips of Zak phase $\theta$.

It is worth noting that there are always two ways to select the centrosymmetric unit cell (i.e., there are two symmetry centers shifted by half of the period $\Delta x= a/2$) which are not equivalent for $\rho$ and $\theta$. When we shift the centrosymmetric unit cell by $a/2$ then $\theta$ flips $0\leftrightarrow\pi$ in every band and $\rho$ is negated in every second gap  (see the Appendix B for explanations).

The more general case is when the unit cells are not selected as centrosymmetric. Then, the symmetry-related criteria for the existence of interface modes cannot be used. However, we can test the continuous transition between two different centrosymmetric selections of the unit cell. We investigate numerically how the shift of the unit cell $\Delta x$, described by the parameter $\delta=\Delta x/a$, influences the existence of SW interface modes.We will keep the centrosymmetric unit cell for the MC on the left side (see Fig.~\ref{fig:unitcell}(b)) and change the selection of the unit cell for MC on the right by sweeping the parameter $\delta$ in the whole range $[0,1)$. For the gradual change of  $\delta$, we should observe the continuous transition of the frequencies of the SW interface modes between the boundaries of the gap.
The SW interface modes for the values $\delta_1$ and $\delta_2$ correspond to the selection centrosymmetric cell for the MC on the right side of the interface. In this case, the observation of interface states must be consistent with the symmetry-related existence conditions for these states.

\subsection{Numerical calculations}
The LLE is solved by Plane Wave Method (PWM), which is suitable for periodic structures. Detailed discussion
of the application of this computational method for planar magnonic crystals is presented in the paper by Krawczyk et al. \cite{Krawczyk2013}. 
Solving LLE with PWM gives us the information of dispersion relation and  SW's eigenmodes.
The bulk properties of SW in single unit cell (infinite MC) are investigated for the one-dimensional unit cell with periodic boundary conditions. The calculations are done in dependence on the bulk parameter: filling fraction ($f\!f$).

To calculate the SW interface modes, we use a supercell approach\cite{RYCHLY18}. We mimic two semi-infinite MC, joined at the interface, by the supercell composed by finite MC of two kinds, each consisting of $N=$ 100 unit cells. 
Inevitable artifact of this approach is existence of second (complementary) interface due to periodic boundary conditions which can also bound the SW interface modes. Therefore, in our calculation we will see two  modes localized on different interfaces. This modes will be degenerated (and will occupy both interfaces at the same frequency) for $\delta=\delta_1$ or $\delta=\delta_2$, where unit cells are centrosymmetric and both interfaces are structurally identical.

For the geometry presented in Fig.~\ref{fig:unitcell}(b), the number of unit cells within each MC should diverge to infinity. However, due to computational power limitations, we are constantly forced to perform the computations on the finite domain. To reproduce the spectrum of SW interface modes satisfactorily, we must consider the large supercell, where the distance between two interfaces is enough to avoid overlapping decaying exponentially "tails" of interface modes. For considered structures, this condition is fulfilled even for the smallest gap (characterized by small decay rates $k_i$) when taking about 100 unit cells of each MCs. Thus length of each 1D MC is $D=Na$, where $N$ is the number of unit cells and $a$ unit cell's width. Such systems can be easily investigated on a desktop computers.

\section{Results}
Firstly, we consider the fictitious planar MC with neglected dipolar interaction.
This step allows us later to isolate the impact of dipolar interaction on the existence conditions of interface modes. 
We assumed a small lattice constant, $a=100$ nm, where dipolar interaction would be negligible anyway. Nevertheless it is important to note, that even for small unit cell system with dipolar interaction would vary in the following aspect: (i) dispersion relation is shifted up; (ii) dispersion relation is dependent on the direction of the external magnetic field; (iii) for $k$ close to 0 group velocity is nonzero; (iv) for $k$ close to 0 precession of magnetization vector is not perfectly circular.

Secondly, we consider the planar MC with included dipolar interaction. To make them meaningful and propose a structure that is accessible experimentally, we assumed lattice constant, $a=1000$ nm. In Appendix \ref{sec:App.A}, we show that the Zak phase for dipolar-exchange system (i.e., the system where the elliptical precession must be taken into account) is defined in the same way as for exchange systems. Therefore, the calculation of the Zak phase and logarithmic derivative can be performed in the same manner.

\subsection{Exchange spin waves}
\begin{figure}[!h]
\begin{centering}
\includegraphics[width=1\columnwidth]{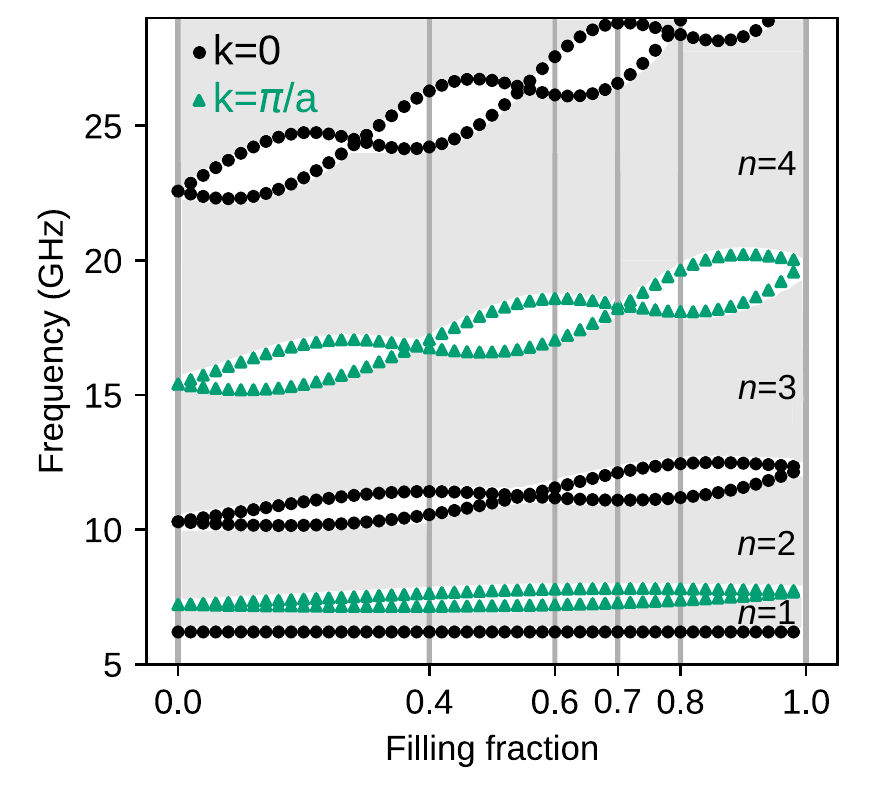}\caption{The evolution of SW spectra in dependence of the bulk parameter: filling fraction $f\!f$ for exchange dominated MC ($a = 100$ nm). White areas represents frequency gaps, while gray regions correspond to the successive frequency bands: $n=1,2,3\ldots$. 
The edges of the bands: $k=0$ and $k=\pi/a$ are marked by black dots and green triangles, respectively.
The vertical lines at $f\!f=0.4,0.7$ and $f\!f=0.6,0.8$ denote the pairs which were interfaced to look for the SW interface modes in common frequency gaps. 
\label{fig:ff_ex}}
\par\end{centering}
\end{figure}

For the system of the small lattice constant $a=100$ nm, we  neglected the dipolar interactions. We start the discussion by analyzing the dependence of the band structure of infinite MC on the bulk parameter -- $f\!f$.
In Fig.~\ref{fig:ff_ex}, the gray and white areas mark the frequency bands and gaps, respectively. For exchange spin waves, the gaps in 1D MC are opened alternatively in the center of the $1^{\rm st}$ Brillouin zone (BZ), i.e., for $k=0$, and edge of the $1^{\rm st}$ BZ, i.e., for $k=\pi/a$ (see also Fig.\ref{fig:dis47}). These values determine the edges of gaps: black dotted line for $k=0$ and green dotted line for $k=\pi/a$.
In the absence of magnetocrystalline anisotropy, the lowest black line is independent of the material or structural parameters, including $f\!f$. Its position reflects the Larmor frequency and is determined only by the external magnetic field.
The edges of higher gaps (i.e., located above successive bands $n=1,2,3,\ldots$) are interwoven and cross each other at specific values of filling fraction $f\!f$ (see also Fig.\ref{fig:dis47}(b)). The values of $f\!f$ at which the gaps are closed correspond to the cases when both the Co and Py layer contains the integer number of the half-wavelengths. It is equivalent to the appearance of standing SWs. The resulting fact is that SW does not scatter on Co|Py interfaces for this particular frequency.
It is worth noting that for $f\!f=0$ or $1$, the system is composed of homogeneous Py or Co where the periodicity and the folding of dispersion relation are introduced artificially. 
The points at which the folded dispersion self-crosses, mark the frequencies for which the (Bragg) gaps start to open when we introduce thin layers of other material. For $f\!f\approx 0$ (for Py matrix with thin Co layer), the gaps are opened at smaller frequencies than the corresponding gaps for $f\!f\approx 1$.
Co has larger exchange length of Co, so the slope of its (parabolic) dispersion relation increases faster with the wavenumber and the sections of dispersion relations (bands) folded into the $1^{\rm st}$ BZ are wider in frequency domain.
Therefore, the gaps not only interwove their edge with increasing $f\!f$ but also push towards higher frequencies. 
Relatively narrow band gaps make designing the magnonic system with neglected dipolar interactions difficult, and proper selection of the system's properties become crucial.

Due to the reversing of the order of the gap's edges, the Zak phases (\ref{eq:Zak}) of two surrounding bands are flipped, and the sign of logarithmic derivative (\ref{eq:rho}) inside the gap is negated. This observation concludes that by adjusting the bulk parameters (i.e., filling fractions $f\!f$) for two 1D MCs joined at the interface, we can adjust the topological parameters of their spectra to obtain a common frequency gap. The matching of boundary conditions expressed by agreement of logarithmic derivatives of the Bloch functions exponentially decaying in the interior of corresponding MC can be achieved (see Appendix \ref{sec:App.C}). 
 
In the numerical studies of interface modes, we will also investigate the more general case, when the unit cell for one MC is not centrosymmetric and thus it is not interfaced with other MC at its symmetry point.

\subsubsection{Interface modes for $f\!f_{L}=0.4$ and $f\!f_{R}=$0.7}
\begin{figure}[ht]
\centering{}\includegraphics[width=1\columnwidth]{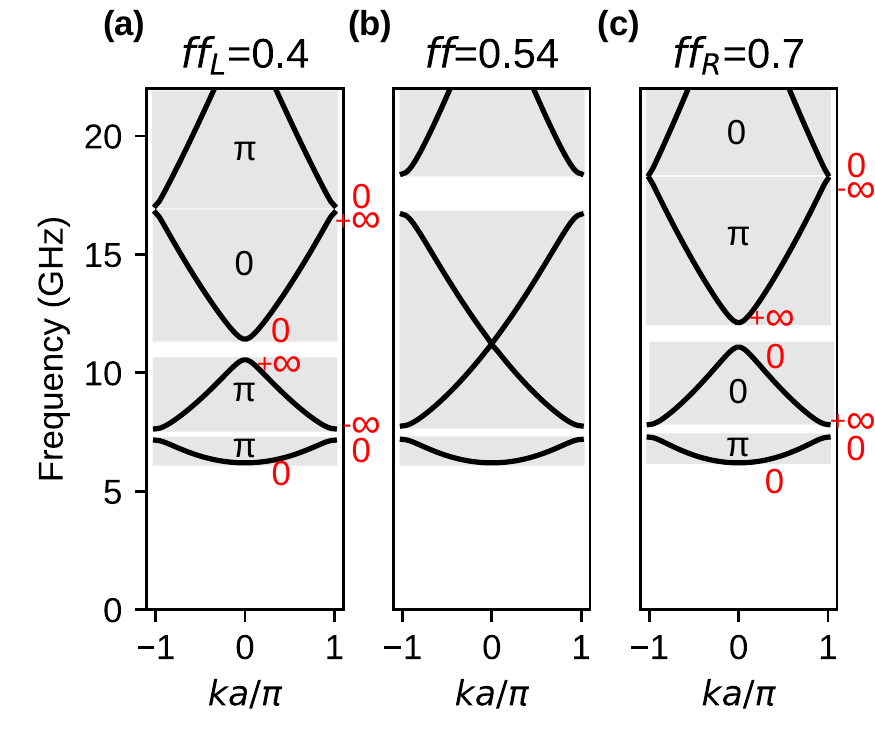}\caption{Closing and reopening of the frequency gap (above the $2^{\rm nd}$ band, in the center of $1^{\rm st}$ Brillouin zone: $k=0$) with the changes of filling fraction $f\!f$ for exchange waves. The successive dispersion relations were plotted when the gap is (a) opened,  $f\!f_L=0.4$, (b) just closed, $f\!f=0.54$, and (c) opened again,  $f\!f_R=0.7$. 0 and $\pi$ labels stand for Zak phase of the band, 0 and $\infty$ (red color) stand for logarithmic derivative on the edges of band gaps. The values of Zak phases and logarithmic derivatives were determined for the case $\delta=\delta_1$ (i.e., for Co strip in the center of the unit cell).
\label{fig:dis47}}
\end{figure}

Fig. \ref{fig:dis47} presents the dispersion relation within the $1^{\rm st}$ BZ for three selected filling fractions $f\!f=0.4,0.54,0.7$. The dispersion branches in Fig. \ref{fig:dis47} are labeled with Zak phase, as well as edges of band gaps with the value of the logarithmic derivative  $\rho$ (see the Appendix \ref{sec:App.B}) -- the marked values are valid for centrosymmetric unit cell, where $\delta=\delta_1$. 
For the value $f\!f\approx0.54$ (Fig.\ref{fig:dis47}(b)), we observe the crossing of second ($n=2$) and third ($n=3$) band and closing the gap between these bands. Due to the band crossing, the Zak phases for $n=2$ and $n=3$ are flipped -- see Fig.~\ref{fig:vec_full}(a, b), where the change of the symmetry of the Bloch mode at the bands' edges are related to the change of Zak phase. 
Now, when we join two semi-infinite MCs of the filling fractions (with centrosymmetric unit cell: $\delta=\delta_1$) $f\!f_{L}=0.4<0.54$ and $f\!f_{R}=0.7>0.54$ (Figs. \ref{fig:dis47}(a, c)), then we can agree with the sign of the logarithmic derivatives of Bloch function on both sides of the interface in the common frequency gap 
(see Fig. \ref{fig:log_drv_04_07}). 
Please note that, for the same sequence of the Zak phases for successive bands, the signs of logarithmic derivatives of Bloch functions decaying to the left or right (i.e., for the left and right MC, respectively) are opposite in corresponding gaps (i.e., in the gaps between the same bands). Therefore, the band crossing is required to negate the sign of logarithmic derivatives ${\rm sign}(\rho)$ in the reopened gap between the crossed bands (see the Appendix \ref{sec:App.B} and \ref{sec:App.C} for more explanations) and ensure the matching of $\rho$ on both sides of the interface.  

\begin{figure}
\begin{centering}
\includegraphics[width=1\columnwidth]{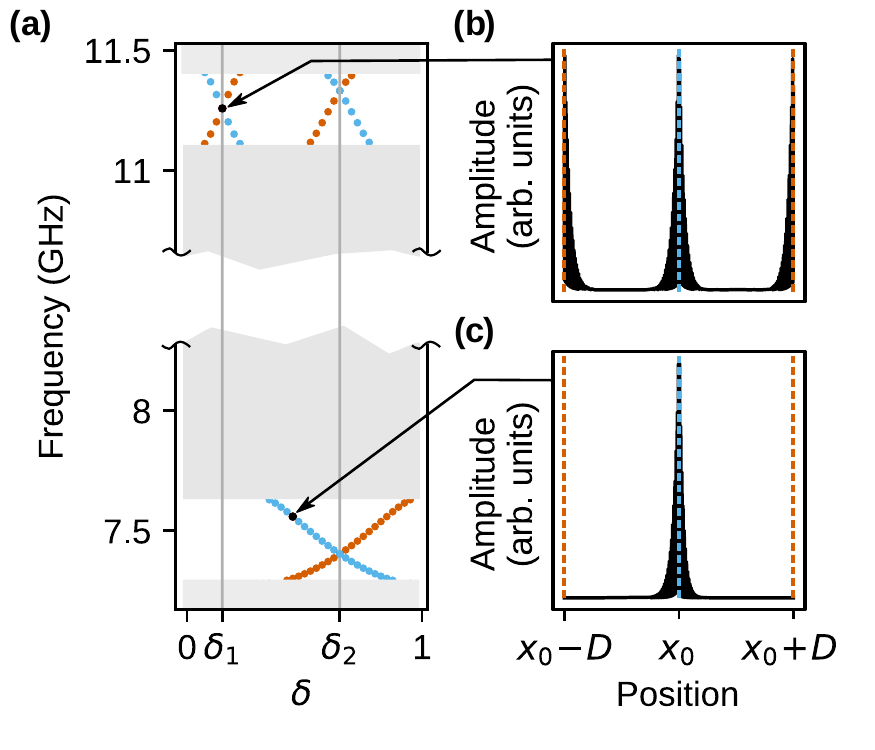}\caption{The SW interface modes localized on the interface of two MCs of $f\!f_{L}=0.4$ and  $f\!f_{R}=0.7$. (a) The spectrum in the function of the surface parameter $\delta$, defining the selection of the unit cells for MC on the right.
The values $\delta_1$ and $\delta_2$ correspond to the centrosymmetric unit cell with Co and Py strip in the middle.
The calculations were performed for supercell approximation, where we considered the final sections of the MCs, each of the size $D=Na$ and composed of $N=100$ unit cells. Due to the application of periodic boundary conditions, we obtain an additional interface between MCs. The frequency of interface mode localized in the center $x=x_0$ (on edge $x=x_0\pm D$) of a supercell is marked by a blue dotted (orange dotted) line.
(b, c) The profiles of the SW interface modes in the gaps above $1^{\rm st}$ and $2^{\rm nd}$ bands. By arrows are indicated the frequency and $\delta$ positions, for which modes are calculated. The interfaces $x=x_0$ and $x=x_0\pm D$ are pointed by blue dashed and orange dashed lines. For centrosymmetric case, mode occupy both interface (b), while for non-centrosymmetric only one (c).
\label{fig:states47}}
\par\end{centering}
\end{figure}

Let us confirm the existence of interface states directly. 
For the two jointed semi-infinite MCs (system presented in Fig.~\ref{fig:unitcell}(b)), we ran PWM calculation (using the supercell approach) and collected SW spectra for wavenumber $k=0$ (i.e., for periodic boundary conditions applied to the supercell). The change of parameter $\delta$, determining the selection of the unit cell in the MC on the right side of the interface, does not influence on the band structure but affect  on the geometry of the interface between two MCs. Therefore, by changing  $\delta$, we influence on the interface modes' frequency without perturbing the band structure. Results are presented in Fig.~\ref{fig:states47}. 
Gray strips indicate frequency ranges where a continuum of states is observed, while white regions represent common frequency gaps. Due to narrow gaps, we broke the frequency axis to extend region of band gaps.
During the evolution of $\delta$, the values: $\delta=\delta_1=0.15$,  $\delta=\delta_2=0.65$, correspond to the scenario, when right MC is build of centrosymmetric unit cell (see inset below fig. \ref{fig:unitcell}(b). 
The calculations of the Zak phase and logarithmic derivative presented in Fig.\ref{fig:dis47} and Fig.\ref{fig:log_drv_04_07} are done for 
$\delta=\delta_1$.

In the common frequency gaps, we find the pairs of interface modes localized on complementary interfaces between both MCs (please see the remarks about the interface in the supercell with periodic boundary conditions in section II.A).
By blue color are marked states occupying central interface and by red color are marked states originating from periodic boundary conditions. For the centrosymmetric cases: $\delta=\delta_{1},\delta_{2}$ (marked by vertical gray lines), both interfaces are equivalent, and the interface states are degenerated and localized on both interfaces at once. By changing the $\delta$ we can tune the frequency of interface modes and traverse the whole range of the frequency gap. For the whole range of $\delta=[0,1)$ (note that $\delta=0$ is equivalent to $\delta=1$) the interface mode traverse the gap even few time, depending on the number of the decaying oscillation of Bloch function in the unit cell. 
These numbers increases for successive gaps. 

\begin{figure}[ht]
\centering{}\includegraphics[width=1\columnwidth]{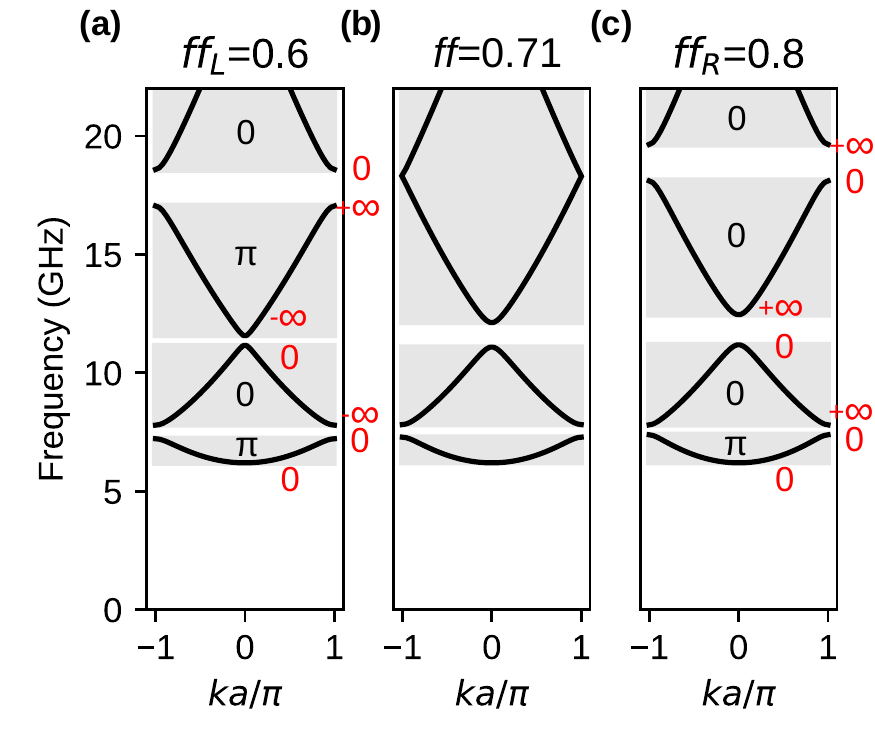}\caption{
{Closing and reopening of the frequency gap (above the $3^{\rm rd}$ band, at the edge of $1^{\rm st}$ Brillouin zone: $k=\pm \pi/a$) with the changes of filling fraction $f\!f$ for exchange waves. The successive dispersion relations were plotted when the gap is (a) opened,  $f\!f=0.6$, (b) just closed, $f\!f=0.71$, and (c) opened again, $f\!f=0.8$. The values 0 and $\pi$ are the Zak phases of the bands, 0 and $\infty$ (red color) stand for logarithmic derivative on the edges of band gaps.  The values of Zak phases and logarithmic derivatives were determined for the case $\delta=\delta_1$ (i.e., for Co strip in the center of the unit cell).}
\label{fig:dis68}}
\end{figure}

\begin{figure}
\centering{}\includegraphics[width=1\columnwidth]{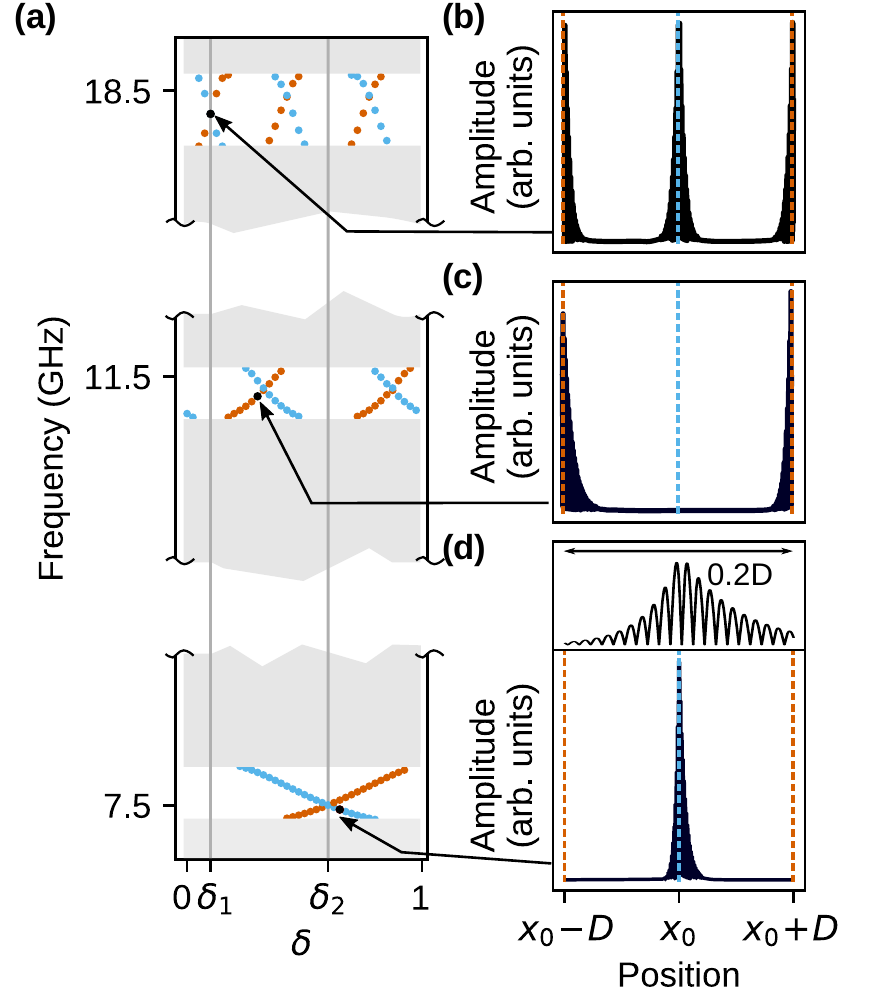}\caption{
\label{fig:states68}
The SW interface modes localized on the interface of two MCs of $f\!f_{L}=0.6$ and  $f\!f_{R}=0.8$. 
(a) The spectrum in the function of the surface parameter $\delta$, defining the selection of the unit cells for MC on the right.
(b-d) The profiles of the SW interface modes in the gaps above $1^{\rm st}$, $2^{\rm nd}$ and $3^{\rm rd}$ bands. We use the same conventions to mark the frequencies and localization of the SW interface modes as in Fig.~\ref{fig:states47}.
Inset above (d) presents a close-up of the SW profile near the interface. It can be seen that despite the fact of decaying, the character of the mode is oscillating.}
\end{figure}

It is worth noting that by the gradual changes of $\delta$, we are continuously transiting between two different selection of centrosymmetric unit cell for the MC on the right side of the interface (corresponding to $\delta_1$ and $\delta_2$). Such change will result in the flipping of the Zak phase for each band ($0\leftrightarrow \pi$) and the negation of logarithmic derivative of Bloch function in every second gap, i.e., for the gaps opened at the edges of $1^{\rm st}$ BZ. 
This effect allows relating the (non)existence of interface modes for the cases: $\delta=\delta_1$ and $\delta=\delta_2$\cite{Zak89}  -- see the Appendix \ref{sec:App.B} for more details. 
The gap around 7.5 GHz (presented in Fig.\ref{fig:states47}(a)) is opened at edge of the $1^{\rm st}$ BZ (between the bands $n=1$ and $n=2$), so the interface modes can exist either for $\delta=\delta_1$ or $\delta=\delta_2$ -- we observe here only for $\delta=\delta_2$. 
Whereas, the gap just above 11 GHz (Fig.\ref{fig:states47}(a)) is opened at the center of the $1^{\rm st}$ BZ (between the bands $n=2$ and $n=3$), which means that the existence conditions of interface modes are the same in both cases ($\delta=\delta_1$, $\delta=\delta_2$) -- here we observe that they exist for both selections of centrosymmetric unit cell. 

\subsubsection{Interface modes for $f\!f_{L}=0.6$ and $f\!f_{R}=$0.8}

The interface modes can also be found when we select different values of filling fractions for MCs on the left and the right side of the interface: $f\!f_{L}=0.6$ and $f\!f_{R}=0.8$ (Fig. \ref{fig:dis68} (a) and (c), respectively).
For the intermediate value of the filling fraction $f\!f=0.71$ (Fig. \ref{fig:dis68}(b)), we observe the crossing of the gap's edges between third ($n=3$) and forth ($n=4$) band -- see Fig.~\ref{fig:ff_ex}. 
The crossing of the bands allows matching the signs of the logarithmic derivatives in the third gap on both sides of the interface between two MCs (see also Supplementary Information, section 2). 
However in Fig. \ref{fig:dis68}(c) we can notice that Zak phase of third ($n=3$) and forth ($n=4$) band is 0. The reason for this is the additional swapping of the Zak phase between the fourth and fifth band gap that is visible in Fig.~\ref{fig:ff_ex}.
Regarding the eq. (\ref{eq:rho_theta}),
the sign of the logarithmic derivative is determined by the Zak phases of all bands below it, so the Zak phase of the band over the gap is irrelevant. The values of the logarithmic derivative at the edges of the gaps and the Zak phases superimposed on Fig.~\ref{fig:dis68}(a-c) was determined for centrosymmetric unit cell, i.e., for $\delta=\delta_1$ -- see also Fig.~2 in Supplementary Information, where the Zak phase was determined from the profiles of the Bloch functions at the edges of the bands.

The results of the SW spectra calculations ($k=0$) of supercell with two jointed MCs for $f\!f_{L}=0.6$ and $f\!f_{R}=0.8$ are presented in Fig.~\ref{fig:states68}. Fig.~\ref{fig:states68}(a) shows the SW spectra in the function of $\delta$.
In the considered frequency range we can see three band gaps between the bands $n=1,2,3,4$. 
The interface states traverse between the edges of gaps, and the number of times increase with the number of band gaps.
Similarly to the previous case ($f\!f_{L}=0.4$, $f\!f_{R}=0.7$), the interface modes appear in pairs and are localized in the middle of the supercell (Fig.~\ref{fig:states68}(d)), at the edge of supercell (Fig.~\ref{fig:states68}(c)) or at both locations (Fig.~\ref{fig:states68}(b)) -- due to degeneracy.  
The spatial oscillation of the interface modes are not visible because of the large number unit cells of each MC ($N=100$) within the supercell. 
However, the inset in top part of Fig.~\ref{fig:states68}(d) shows the zoomed profile of interface mode in the vicinity of the interface. This modes has one oscillations per unit cell, therefore its logarithmic derivative flips its sign once as the interface $x=x_0$ appears at the successive locations in Co and Py layers (as $\delta$ ranges from 0 to 1). As a results, this mode traverses once across the gap in the whole range of $\delta$.

We can analyze the existence of interface modes for two centrosymmetric cases $\delta=\delta_1$ and $\delta=\delta_2$ in similar way as for the structure  $f\!f_{L}=0.4$ and $f\!f_{R}=0.7$. The gaps around 7.5 and 18.5 GHz open at the edge of the $1^{\rm st}$ BZ, therefore the interface state can be observed only for one selection of centrosymmetric unit cell. For the gap opened around 11.5 GHz, the absence of interface modes for $\delta=\delta_1$ implies the nonexistence of these states at $\delta=\delta_2$.

\subsection{Dipolar-exchange spin waves}
\begin{figure}[ht]
\begin{centering}
\includegraphics[width=1\columnwidth]{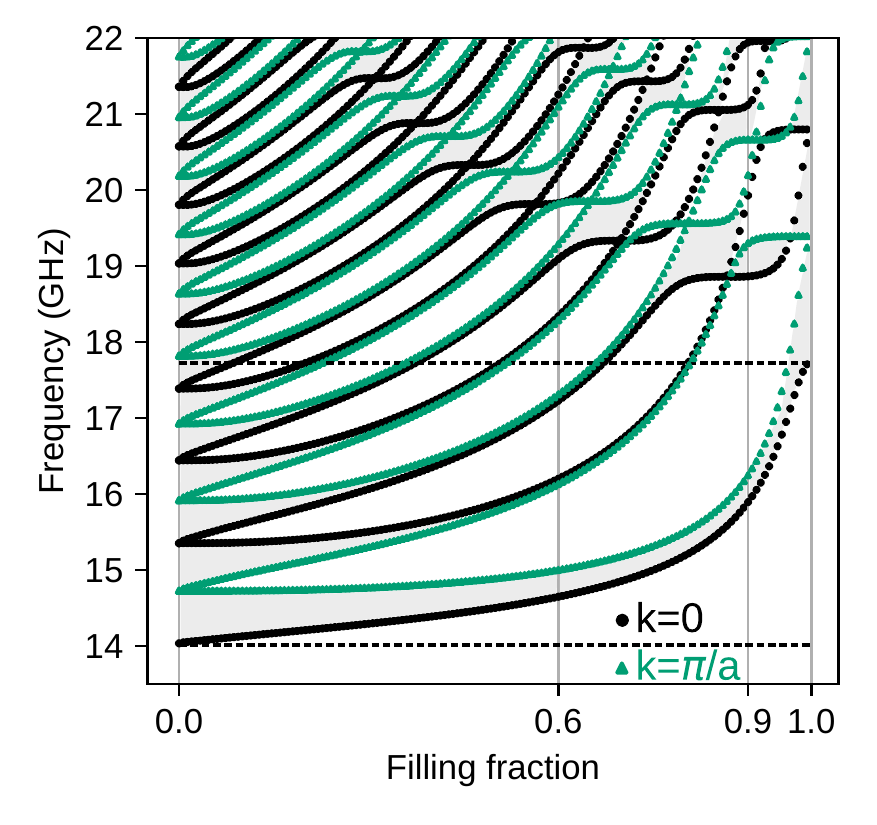}\caption{
The evolution of dipolar-exchange SW spectra in dependence on the bulk parameter: filling fraction $f\!f$ ($a = 1000$ nm). The vertical lines at  $f\!f=0.6,0.9$ denote the pair which were interfaced to look for the SW interface modes in common frequency gaps. We used the same convention to mark frequency bands, gaps and their edges as in Fig.~\ref{fig:ff_ex}. The higher (lower) horizontal dashed line marks the FMR frequency for uniform Co (Py) layer.
\label{fig:ff_dip}}
\par\end{centering}
\end{figure}

Let us now present the results for the system with included dipolar interactions. To observe the impact of dipolar interactions, we expanded the sizes of the system. The lattice constant is now larger by one order of magnitude: $a=1000$ nm, referring to discussed case with exchange waves. Like in the previous section, we start from the analysis of SW spectra. Its dependence on the filling fraction $f\!f$ is presented in Fig.~\ref{fig:ff_dip}. The first observation is that the fundamental mode is sensitive to magnetic parameters, contrary to spectrum with active only exchange interactions. The bottom of the $1^{\rm st}$  band ($k=0$) is strongly dependent on $f\!f$. Starting from $f\!f=0$, the frequency slowly increases from 14 GHz, while around $f\!f=0.8$ it rises quickly, reaching ultimately about 17.5 GHz for $f\!f=1$. The frequencies of the lowest SW modes (for $k=0$) are just the frequencies of the ferromagnetic resonance (FMR), which is expressed as $f=\frac{|\gamma|\mu_0}{2\pi}\sqrt{H_0(H_0+M_S)}$ for homogeneous layer of Py ($f\!f=0$) or Co ($f\!f=1$), differing significantly in saturation magnetization $M_{S,Py}<M_{S,Co}$. Therefore, the FMR frequency increases  with the increase of filling fraction: $0<f\!f<1$. We marked the FMR frequencies of uniform Py and Co films by dashed lines.

The frequencies for other edges of the bands/gaps quickly increase with the $f\!f$ too. However, their interweaving is not observed for low values of $f\!f$ and low frequencies. It can be understood when we notice that the band crossing in 1D bi-component magnonic crystal requires the oscillatory solution in both components (strips, layers). In our system, the evanescent waves exist in Co strips for the frequencies below the FMR frequency of uniform Co. 
Additionally, due to confinement effect the frequencies of oscillatory modes in Co are increased for smaller $f\!f$ where the Co strips are narrow.
Therefore, to find the interface states, we selected a pair of MCs of relatively high filling fractions $f\!f$: 0.6|0.9 for which the edges of gaps can have different number of crossing points, that is related to different topological properties of their band structures.

\begin{figure}[ht]
\begin{centering}
\includegraphics[width=1\columnwidth]{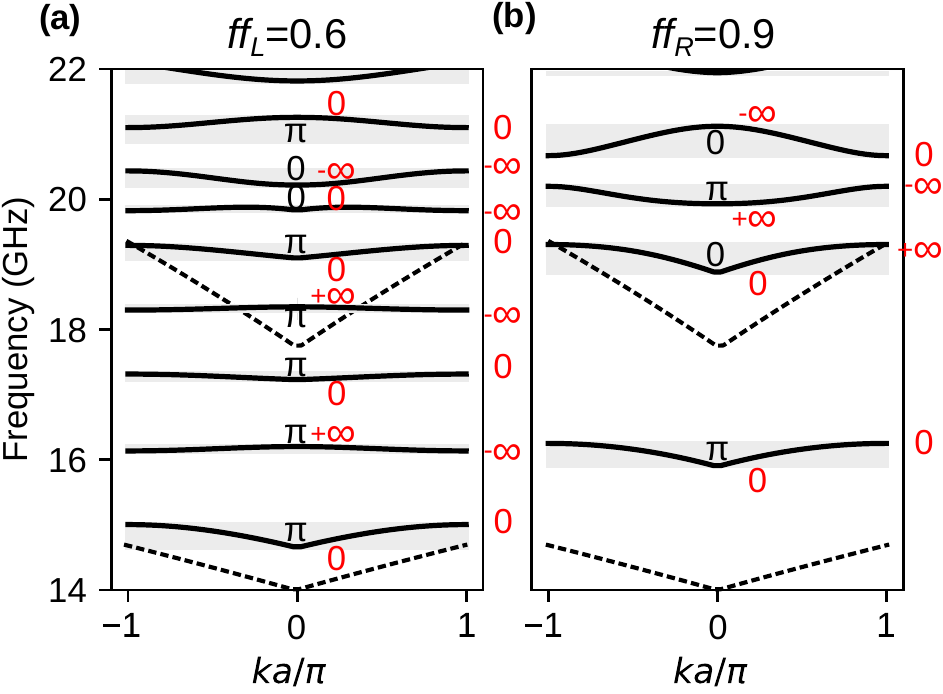}\caption{
The dipolar-exchange SW dispersion relations for two MCs which differ only in filing fractions: (a) $f\!f_{L}=0.6$, (b) $f\!f_{R}=0.9$. The spectrum for $f\!f=0.9$ is shifted up in the frequency scale due to dipolar interaction, and the successive gaps in both spectra do not match each other.
The forbidden ranges for interfaced MCs ($f\!f_{L}=0.6$ and $f\!f_{R}=0.9$) can originate from overlapping of various frequency gaps in both MCs -- see Fig.~\ref{fig:ff_dip}.  The values 0 and $\pi$ are the Zak phases of the bands. 0 and $\infty$  (red color) stand for logarithmic derivative on the edges of bad gaps.
Dashed lines represent dispersion relation for a uniform system made from Py (branch starts at ~14GHz) and made from Co (branch starts at ~17.5 GHz).
\label{fig:disp_dip}}
\par\end{centering}
\end{figure}

\begin{figure}[ht]
\begin{centering}
\includegraphics[width=1\columnwidth]{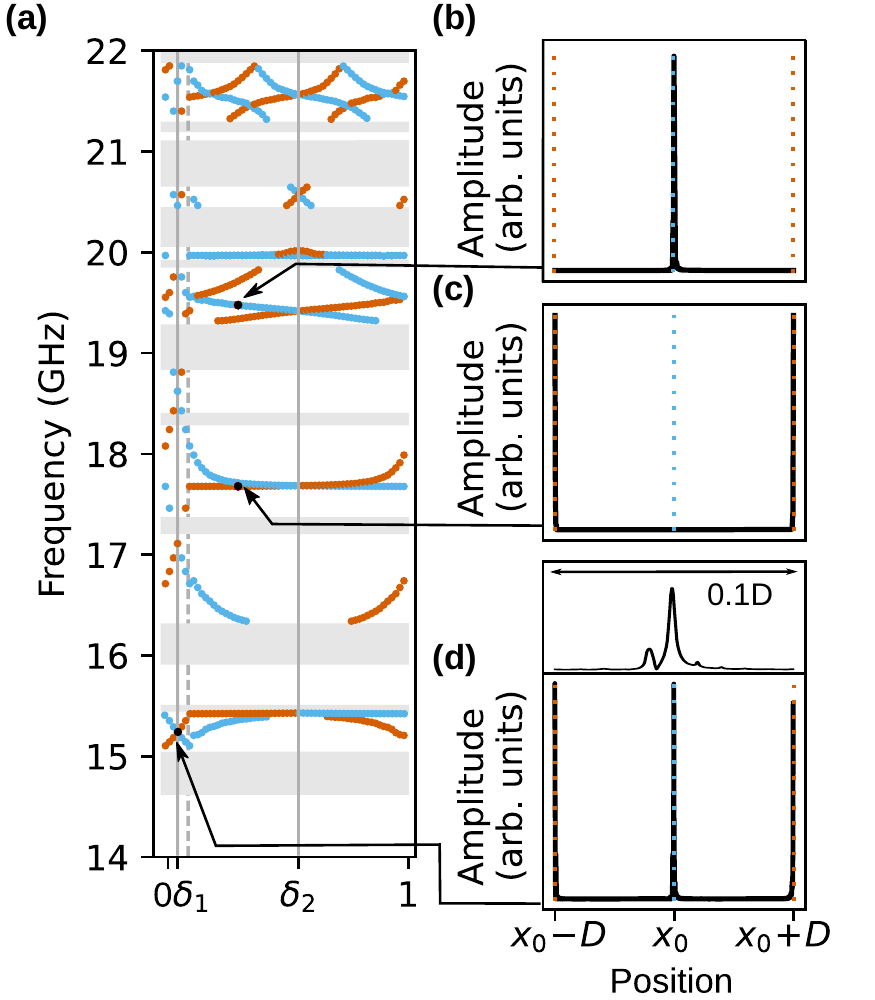}\caption{
The dipolar-exchange SW interface modes localized on the interface of two MCs of $f\!f_{L}=0.6$ and  $f\!f_{R}=0.9$.
(a) The spectrum in dependence on the surface parameter $\delta$. We use the same conventions to mark the frequencies and localization of the SW interface modes as in Fig.~\ref{fig:states47}.
The gray dashed line in (a) denotes $\delta=1-f\!f$ where the Py|Co interface is on the left edge of the unit cell.
(b-d) The profiles of the SW interface modes in common (from both MCs) forbidden frequency ranges. Above the (d) is presented close-up to the $x=x_0$. SW decay exponentially in Co strips and have one oscillation per Py strip.
\label{fig:dys_mod_dip}}
\par\end{centering}
\end{figure}

The dispersion relations for these MCs are presented in Fig.~\ref{fig:disp_dip}.
For reference, we showed the dispersion relations of the uniform film made of Co and Py in the range of ($-\pi/a$, $\pi/a$). They are marked with dashed lines.
Thanks to this comparison, we can attribute the first branch of MC in Fig. \ref{fig:disp_dip}(a) as excitation in Py, while the fifth in Co.
In Fig. \ref{fig:disp_dip}(b), the first branch of MC is excitation in Py, while the third one is excitation in Co. 
The character of SW's profile is implicitly presented in Appendix \ref{sec:App.D}. 
These bands are shifted up with respect to fundamental excitation in the uniform film due to the confinement and dipolar pinning in the strips \cite{Centala19}. SW is forced to be quantized within the strip, and due to this, its frequency increase. For $f\!f=0.9$ and the Py strip, this effect is the most significant because the Py strip is the narrowest -- it has there only 100 nm.
For $f\!f_{R}=0.9$, two first bands have a minimum at the center of BZ, and a maximum at the edge of BZ. It violates the typical scenario when the maximum and minimum of two successive bands appear at the same position in BZ, i.e., the case when we observe only the direct gaps. 
This peculiarity of the system can be explained when we notice that the second band is not a result of folding the first band into the $1^{\rm st}$ BZ, but is related to the lowest oscillatory solution in the Co strips. In the first band, we observe in Co the evanescent excitation forced by the magnetization dynamics in Py. 
This consideration is supported by spatial profiles of the modes, attached to Appendix \ref{sec:App.D}.
It is worth noting that the lowest mode in Co can be also identified by the non-zero group velocity, characteristic of dipolar dynamical coupling in the limit $k\rightarrow 0$. 

The spectrum of considered 1D MC's is very rich and is dramatically modified with the change of the filling fraction -- the bands' positions and their separations (width of the gaps) vary significantly with $f\!f$. Unlike the system with exchange interaction only, the positions of corresponding band gaps (i.e., the band of the same index $n$) are different. For example, the first band gap for $f\!f_{R}=0.9$ is located in the frequency range corresponding to second and third band gaps for $f\!f_{L}=0.6$. The bands are relatively narrow which proves that we are operating in crossover dipolar-exchange regime. The flat dispersion branches in Fig.~\ref{fig:disp_dip} means the low value of the group velocity for SW. 

In Fig.~\ref{fig:disp_dip}, we marked the Zak phases for centrosymmetric unit cells ($\delta=\delta_1$). The values were determined by the inspection of the profiles of the Bloch functions at the edges of the bands (see Fig.~\ref{fig:vec_full}(e,f)). 
The sequences of Zak phases allows determining the sign of logarithmic derivatives of Bloch function, $\rho$, in frequency gaps for MCs at left and right side of the interface. The values of $\rho$ at edges of bands are marked in Fig.~\ref{fig:disp_dip} with red color. 

The careful inspection of the sign of $\rho$ in common frequency gaps of both MCs (marked as white areas in Fig.~\ref{fig:disp_dip}) allows indicating three common gaps in which the signs of $\rho$ are opposite: gap below $1^{\rm st}$ band, gap around 18.5 GHz and tiny gap, slightly above 21 GHz. 
In these gaps, we did not find numerically any interface modes: Fig. \ref{fig:dys_mod_dip} does not present solution at these frequencies for $\delta=\delta_1$.
Concluding, analysis of logarithmic derivative for dipolar-exchange waves is also valid tool to describe the existence criteria for interface modes.

This system posses some peculiarities. Due to disturbing the typical sequence of dispersion branches, logarithmic derivative can take the same values on the edges of band gaps. For $f\!f_{R}=0.9$ first band gap $\rho$ is equal to 0 on both of the edges, and in second gap is equal to $-\infty$.
We were also not able to plot the logarithmic derivatives of Bloch functions, as we did for exchange waves in Appendix \ref{sec:App.C}. It is related to the difficulties with the determination of the complex wavenumber in the indirect gaps, needed to be specified to run the PWM calculations. 

For selected pair of the filling fractions ($f\!f_{L}=0.6$,  $f\!f_{R}=0.9$) of two MCs, we performed the numerical studies of the modes localized at interface between MCs. Using the PWM and supercell approach we calculated the frequencies and profiles of interface modes localized on two complementary interfaces. The frequencies were plotted in dependence on the parameter $\delta$ which describes the selection of the unit cell in the MC on the right side of the interface. Fig.~\ref{fig:dys_mod_dip}(a) presents the evolution of interface states in the function of $\delta$. The first observation is that the dependencies of the frequencies of interface modes on the unit cell shift $\delta$ can change their slopes abruptly. This effect is related to the sensitivity of dipolar interaction on the interfaces. With an increasing value of $\delta$, the narrow Py strip ($f\!f=0.9$) moves into the center of the unit cell  (see the inset below Fig.~\ref{fig:unitcell}(b) for graphical illustration). Till the $\delta$ reach the value marked by the gray dashed line $\delta=1-f\!f$ , the Py strip on the interface is widening. After that, the Co strip starts to be located on the interface, so the MC on the left starts to be interfaced with different material on the right side.

The another interesting finding is the absence of multiple traverses of the interface modes across the common gap for the swap of parameter $\delta$, observed  for small frequencies (below ~19.5 GHz, where the modes start spatially oscillate in Co) and for the larger values of the $\delta$ ($1-f\!f<\delta<1$, where the left edge of the unit cell of the right MC appears in Co strip). In this range of the $\delta$, the solution at the edge of unit cell (which is the right side on interface between MCs, as well) cannot change the sign. It is because of the evanescent profile of the mode in Co. This excludes multiple flips of the sign of the mode at the interface while $\delta$ is changed.

Fig.~\ref{fig:dys_mod_dip}(b-d) shows the profile of the interface states for selected values of $\delta$ and frequency. They are strongly localized on the interface. Inset above (d) presents a close-up of the interface region. The oscillatory character is only in the Py strips, while in Co, the amplitude decay exponentially. The strong localization is a consequence of wide gaps in which the imaginary parts of wavevector (describing the exponential rate of localization) can reach large values.

\section{summary}
We have presented a comprehensive study on the existence of interface SW states in 1D planar magnonic crystals, using a continuous model of magnetization dynamics for the exchange and dipolar-exchange waves.

We have related bulk parameter in magnonic crystal to the symmetry-related conditions of existence of interface states: (i) the concept of Zak phase, which is a topological characteristic of individual bands in the frequency spectrum was connected to (ii) the logarithmic derivative of the Bloch function on both sides of the interface, expressing the boundary conditions for interface modes in the band gaps. We have also performed numerical results that allowed us to consider the behavior of the interface modes for non-centrosymmetric unit cells. We have shown that this degree of freedom can be used to induce or vanish the interface state in desired band gap.

Full analogy to the already investigated electronic and photonic systems is observed in the magnonic system where the dipolar interactions are neglected. 
For the dipolar-exchange waves, however, the analysis becomes more complex. 
We have observed new effects specific to dipolar interaction: (i) rarer crossings of band gap edges -- the band gaps do not close in a wide range of the filling fraction and the selection of pair of MCs with band structures supporting interface modes is challenging; (ii) in the lower-frequency range (i.e., in lower band gaps) the observed interface modes do not traverse the band gap edges with shifting MC unit cell.
Nevertheless, we have found numerous interface modes, and their existence (for centrosymmetric unit cell) was determined from the symmetry criterion of the Bloch function on the band edges.

\section{Acknowledgements}
The authors thank Maciej Krawczyk for useful comments and fruitful discussion. S.~M. and J.~W.~K. would like to acknowledge the financial support from the National Science Centre, Poland, pojects No. UMO-2020/36/T/ST3/00542 and No. UMO-2020/37/B/ST3/03936. 
This study was partially supported by the Polish agency NAWA under the grant No PPN/BUA/2019/1/00114.
J.~W.~K. thanks for the support form the Foundation of Alfried Krupp Kolleg Greifswald.
\newpage
\appendix

\section{Zak phase for spin waves in planar magnonic crystals}
\label{sec:App.A}
\subsection{Exchange spin waves}
The linearized Landau-Lifshitz-Gilbert equation can be written in the following form when the dipolar interactions are neglected $\boldsymbol{H}_{\rm dm}=0$:

\begin{equation}
\begin{cases}
\partial_{t}m_{\parallel}\left(x,t\right)=|\gamma|\mu_{0}\left(\hat{L}(x)m_{\perp}\left(x,t\right)-V m_{\perp}(x,t)\right)\\
\partial_{t}m_{\perp}\left(x,t\right)=|\gamma|\mu_{0}\left(-\hat{L}(x)m_{\parallel}\left(x,t\right)+V m_{\parallel}(x,t)\right),
\end{cases}\label{eq:LLlin}
\end{equation}
where the function $V(x)$ and operator $\hat{L}(x)$ are defined as:
\begin{equation}
\begin{cases}
V(x)=\left(\partial_{x}\lambda^{2}\partial_{x}M_{\rm S}(x)\right)+H_{0}\approx H_0\\
\hat{L}(x)=M_{\rm S}(x)\partial_{x}\lambda^{2}(x)\partial_{x},
\end{cases}\label{eq:VL}
\end{equation}
where $\lambda$ is exchange length. We neglect the static term $\partial_{x}\lambda^{2}\partial_{x}M_{\rm S}(x)$ because it is nonzero only at the interface between strips and can be neglected in numerical computations\cite{Krawczyk2013}. 
To find the eigenmodes, we are considering the harmonic dynamics in time: $\boldsymbol{m}\left(x,t\right)=\boldsymbol{m}(x)e^{i\omega t}$. The equation (\ref{eq:LLlin}) is a Sturm-Liouville problem, 
can be written in the form analogous to Schr\"odinger equation:
\begin{equation}
\partial_{t}\left|m\right>=\varLambda\left|m\right>,\label{eq:SchEq}
\end{equation}
where we used the notation $\left|m\right>:=[m_\parallel(x),m_\perp(x)]^{T}e^{i\omega t}$ and the matrix $\varLambda$ is defined as following:
\begin{equation}
\varLambda=|\gamma|\mu_{0}\left(\begin{array}{cc}
0 & -V+\hat{L}\\
V-\hat{L} & 0
\end{array}\right).
\end{equation}

Let us consider the continuous transition of the vector $\left|m\right>$  in the momentum space of the wavenumber $k$, after which it acquires the  phase $\varphi(t)$:
\begin{equation}
\left|m'\right>=e^{i\varphi}\left|m\right>.
\end{equation}
Eq. (\ref{eq:SchEq}) is satisfied for $\left|m'\right>$ as well.  We can write the following relation resulting from (\ref{eq:SchEq}):
\begin{equation}
   \left<m\middle|\partial_{t}\middle|m'\right>=\left<m\middle|\varLambda\middle|m'\right>,\label{eq:zak_deri} 
\end{equation}
when we define the inner product $\left<f_1\right.\left|\,f_2\right>:=\int_{-a/2}^{a/2}\left(f_{1,\parallel}^*(x)f_{2,\parallel}(x)+f_{1,\perp}^*(x)f_{2,\perp}(x)\right)dx$. By differentiating $\left|m'\right>$ in time and using the identity: $\partial_k=\dot{k}\partial_t$, we can write (\ref{eq:zak_deri}) in the form:

\begin{equation}
i\dot{\varphi}\left<m\middle|m\right>+\left<m\middle|\partial_k\middle|m\right>\dot{k}=i\omega\left<m\middle|m\right>.\label{eq:zak_deri2}
\end{equation}
Taking in to account that $\left|u\right>=\left|m\right>e^{-ikx}e^{-i\omega t}$, we obtain form (\ref{eq:zak_deri2}):
\begin{equation}
  \varphi=\underbrace{\int_{o}^{t}\omega\left(t\right)dt}_{\varphi_t}+\underbrace{\int_{k\left(0\right)}^{k\left(t\right)}\frac{i\left<u\middle|\partial_{k}\middle|u\right>}{\left<u\middle|u\right>}dk}_{\varphi_g}.
\end{equation}
The phase $\varphi$ contains an additional term $\varphi_g$ which is distinguishable from the phase  $\varphi_t$ acquired from the temporal evolution of the eigenmode $\left|m\right>$ . The geometrical phase $\varphi_g$ collected when Bloch function $\left|m\right>$ passes the periodic path in the space of $k$-number (i.e., when $k$ is real and ranges from -$\pi/a$ to $\pi/a$) is called Zak phase:
\begin{equation}
   \theta=\int_{-\pi/a}^{\pi/a}\frac{\Im\left<u\middle|\partial_{k}\middle|u\right>}{\;\;\left<u\middle|u\right>}dk.\label{eq:zak_ph_ex}
\end{equation}

\subsection{Dipolar-exchange spin waves}
The demagnetizing field $\boldsymbol{H}_{\rm dm}(x,z,t)=-\nabla \varphi(x,z,t)$ is calculated under magnetostatic approximation by finding the magnetostatic potential $\varphi(x,z,t)$ from the Gauss equation. For the layer which is periodically modulated in the plane, the demagnetizing field was calculated using the method proposed by J. Kaczer\cite{Kaczer1974}, which is based on the Fourier expansion. For thin planar structure, the demagnetizing field does not change significantly inside the magnetic layer and we  took its value in the middle of the layer ($z=0$) as a representative for the whole cross-section of the layer. The dynamic demagnetizing field is expressed in terms of the coefficients of the Fourier expansion of the Bloch function $\left|m\right>$, therefore the Eq. (\ref{eq:SchEq}) must be written in the Fourier space as well:
\begin{equation}
\partial_{t}\left|\tilde{m}\right>=\tilde{\varLambda}\left|\tilde{m}\right>\label{eq:SchEqF}
\end{equation}
where $\left|\tilde{m}\right>:=[u_{\parallel,G_0},\ldots,u_{\parallel,G_{n}},\ldots,u_{\perp,G_0},\ldots,u_{\perp,G_{n}},\ldots]^{T}$ $\times e^{i k x}e^{i\omega t}$ and matrix $\tilde{\varLambda}$ takes a form:
\begin{equation}
\tilde{\varLambda}=|\gamma|\mu_{0}\left(\begin{array}{cc}
0 & -\tilde{V}+\tilde{L}+\tilde{D}^{\parallel,\perp}\\
\tilde{V}-\tilde{L}-\tilde{D}^{\perp,\parallel} & 0
\end{array}\right).
\end{equation}
The matrices $\tilde{L}_{i,j}$ and $\tilde{V}_{i,j}$ are related to (\ref{eq:VL}) and describes the impact of external field and exchange interactions, respectively. The matrices $\tilde{D}^{\perp,\parallel}$ and $\tilde{D}^{\parallel,\perp}$ describe the dynamic dipolar interactions in 1D planar magnonic crystal\cite{Kaczer1974, Krawczyk2013} -- the difference between them is reflected in the ellipticity of pressecion of dipolar spin waves. The explicit form of these matrices is:
\begin{eqnarray}
    \tilde{L}_{i,j}&=&-\sum_l(k+G_i)(k+G_l)\lambda^2_{G_l-G_j}M_{S,G_i-G_l},\nonumber\\
    \tilde{V}_{i,j}&=&H_0\delta_{i,j},\nonumber\\
    \tilde{D}^{\perp,\parallel,}_{i,j}&=&-\left(1-e^{-|k+G_j|d/2}\right)M_{S,G_i-G_j},\nonumber\\
     \tilde{D}^{\parallel,\perp}_{i,j}&=&-e^{-|k+G_j|d/2}M_{S,G_i-G_j},\label{eq:frourier}
\end{eqnarray}
where $d$ is thickness of the layer. The material parameters, i.e., saturation magnetization $M_{S}(x)$, exchange length $\lambda(x)$ and the components of Bloch function: $m_\parallel(x)$, $m_\perp(x)$ are expanded into Fourier series:
\begin{eqnarray}
    M_{S}(x)&=&\sum_{n=0} M_{S,G_n}e^{i G_n x},\nonumber\\
    \lambda(x)&=&\sum_{n=0} \lambda_{G_i}e^{i G_n x},\nonumber\\
    m_{\parallel(\perp)}(x)&=&\sum_{n=0} u_{\parallel(\perp),G_n}e^{i(G_n+k)x}.\label{eq:FT}
\end{eqnarray}
The set $\{G_n\}=0,\pm 2\pi/a,\pm 4\pi/a,\ldots,\pm n 2 \pi/a,\ldots$ denotes the reciprocal lattice numbers. 

To prove that the formula (\ref{eq:zak_ph_ex}) also applies to dipolar-exchange waves, we need to show that $\left<u\middle|\partial_{k}\middle|u\right>=\left<\tilde{u}\middle|\partial_{k}\middle|\tilde{u}\right>$ and $\left<u \middle|u\right>=\left<\tilde{u}\middle|\tilde{u}\right>$, where $\left|\tilde{u}\right>=\left|\tilde{m}\right>e^{-ikx}e^{-i\omega t}$:
\begin{eqnarray}
   &&\left<u\middle|\partial_{k}\middle|u\right>= \nonumber\\
   &&\int_{-a/2}^{a/2}\left(u_{\parallel}^*(x)\partial_k u_{\parallel}(x)+u_{\perp}^*(x)\partial_k u_{\perp}(x)\right)dx=\nonumber\\
    &&\sum_{i,j}\!\!\underbrace{\int_{-a/2}^{a/2}\!\!e^{i(G_i-G_j)x}dx}_{\delta_{i,j}}\left(u_{\parallel,G_i}^*\partial_k u_{\parallel,G_j}+u_{\perp,G_i}^*\partial_k u_{\perp,G_j}\right)=\nonumber\\
    &&\sum_{i}\left(u_{\parallel,G_i}^*\partial_k u_{\parallel,G_i}+u_{\perp,G_i}^*\partial_k u_{\perp,G_i}\right)=:\left<\tilde{u}\middle|\partial_{k}\middle|\tilde{u}\right>.\label{eq:inprod}
\end{eqnarray}
The relation $\left<u \middle|u\right>=\left<\tilde{u}\middle|\tilde{u}\right>$ can be proven in the same way. Starting from (\ref{eq:SchEqF}), we can then show that the Zak phase for dipolar-exchange waves is also equal:
\begin{equation}
   \theta=\int_{-\pi/a}^{\pi/a}\frac{\Im\left<\tilde{u}\middle|\partial_{k}\middle|\tilde{u}\right>}{\;\;\left<\tilde{u}\middle|\tilde{u}\right>}dk=\int_{-\pi/a}^{\pi/a}\frac{\Im\left<u\middle|\partial_{k}\middle|u\right>}{\;\;\left<u\middle|u\right>}dk.\label{eq:zak_ph_dip}
\end{equation}
In (\ref{eq:FT}, \ref{eq:inprod}) we omitted the indexing of $u_{\parallel(\perp)}(x)$ and their Fourier coefficients $u_{\parallel(\perp),G_n}$ by the band number and did not marked their dependence on the wavenumber $k$.  

\section{Zak phase and logarithmic derivative for the crystal with centrosymmetric unit cells}
\label{sec:App.B}
For 1D crystal the Bloch function at $k=0$ or $k=\pm \pi/a$  is periodic ($\boldsymbol{m}(x+a)=\boldsymbol{m}(x)$) or anti-periodic ($\boldsymbol{m}(x+a)=-\boldsymbol{m}(x)$), respectively. Moreover, for the crystals with centrosymmetric unit cells, the Bloch function are even ($\boldsymbol{m}(x_s+x)=\boldsymbol{m}(x_s-x)\Rightarrow\left.\partial_x\boldsymbol{m}=0\right|_{x=x_s}$) or odd ($\boldsymbol{m}(x_s+x)=-\boldsymbol{m}(x_s-x)\Rightarrow\left.\boldsymbol{m}=0\right|_{x=x_s}$) function, with respect to each of two symmetry centers $x_s=na$ or $x_s=a/2+na$, and can be normalized to be real-valued. This gives four possible type of bands, by considering the (even or odd) symmetry of the Bloch function on each of two edges of the band. It is worth noting, that the symmetry of Bloch function is the same at both symmetry centers ($x_s=0+na$ and $x_s=a/2+na$) only for $k=0$ whereas it is reversed (from even to odd or from odd to even) when we change the symmetry point, for $k=\pi/a$. Therefore, it is better to use Wannier functions, defined as:
\begin{equation}
  \boldsymbol{a}(x-na)=\sqrt{\frac{a}{2\pi}}\int_{-\pi/a}^{\pi/a}\boldsymbol{m}_{k}(x)e^{-ikna}dk,
\end{equation}
to classify the symmetry of the bands. The Wannier functions characterize whole band (do not depend on the wavenumber $k$). For the case of crystal of centrosymmetric unit cells, they are exponentially localized around one of symmetry centers ($x_c=0$ and $x_c=a/2$), and are either even or odd with respect to this symmetry center. The periodicity (and anti-periodicity)  and symmetry of the Bloch function at $k=0$ ($k=\pi/a$) can be strictly connected to the properties of the Wannier functions by the relation:
\begin{equation}
    \boldsymbol{m}_k(x)=\sqrt{\frac{a}{2\pi}}\sum_{n=-\infty}^{\infty}\boldsymbol{a}(x-na)e^{ikna},\label{eq:bloch_wannier}
\end{equation}
which can be used to express the Zak phase in terms of Wannier functions. Assuming that the Bloch functions are normalized $\left<m \middle|m\right>=2\pi/a\Leftrightarrow \left<a \middle|a\right>=2\pi/a$, we can obtain from (\ref{eq:zak_ph_dip}) and (\ref{eq:bloch_wannier}):
\begin{eqnarray}
    \theta&=&\frac{2\pi}{a}\int_{-\pi/a}^{\pi/a}i\left<u\middle|\partial_{k}\middle|u\right>dk\nonumber\\&=&\int_{-a/2}^{a/2}\Big(\sum_{n,n'=-\infty}^{\infty}\!\!(x-na)\,\boldsymbol{a}^{*}(x-n'a)\cdot\boldsymbol{a}(x-na).
    \nonumber\\&&\times\underbrace{\int_{-\pi/a}^{\pi/a}e^{ik(n-n')a}dk}_{2\pi/a\,\delta_{n,n'}}\Big)dx
    \nonumber\\&=&\frac{2\pi}{a}\int_{-\infty}^{\infty}x|\boldsymbol{a}(x)|^2dx\nonumber\\&=&\frac{2\pi}{a}\int_{-\infty}^{\infty}x(|a_{\parallel}(x)|^2+|a_{\perp}(x)|^2)dx,\label{eq:Zak_Wannier}
\end{eqnarray}
where $a_{\parallel}(x)$ and $a_{\perp}(x)$ denote the components of Wannier function corresponding to in-plane and out-of-plane components of Bloch function. For the system with centrosymmetric unit cells, the integral $\theta=\frac{2\pi}{a}\int_{-\infty}^{\infty}x|\boldsymbol{a}(x)|^2dx$ takes only two possible values $0$ and $\pi$ which correspond to different symmetry center: ($x_c=0$ or $x_c=a/2$) at which the Wannier function, related to given band, is localized. This allows spliting all band to two disjoint topological classes where the Zak phase $\theta$ is equal to $0$ or $\pi$. Please note that in general case, i.e., when the unit cells are not centrosymmetric then the Zak phase can take arbitrary value. To prove the quantized values of Zak phase for centrosymmetric unit cell, we need to discuss the symmetry of function $x|\boldsymbol{a}(x)|^2$. This function is odd when $x_c=0$, regardless if $\boldsymbol{a}(x)$ is even or odd, which gives $\theta=0$. For $x_c=a/2$, we need make a substitution $x\rightarrow x+a/2$ for the variable inside the integral. Then, we can find that the expression can be split into two terms: $(x+a/2)|\boldsymbol{a}(x+a/2)|^2=x|\boldsymbol{a}(x+a/2)|^2+a/2|\boldsymbol{a}(x+a/2)|^2$, where the first one is odd and second one is even and does not vanish after integration, which gives $\theta=\pi$. The above discussion relates the symmetry of the Bloch function on the edges of the band to the Zak phase for this band in the structures with centrosymmetric unit cells. {\em If the symmetry of Bloch function, respect to the center of unit cell, is the same on both edges of the band then the Zak phase for this band is equal to $0$, otherwise to $\pi$.} As we noticed at beginning of Appendix B, the centering of the unit cell at alternative symmetry center $x_s$ (i.e., shifting it by $a/2$) changes the symmetry of the Bloch function on one edge of the band only, i.e., for the edge at which $k=\pi/a$. As a result {\em the shift of the centrosymmetric unit cell by the half of the period: $x_s\rightarrow x_s+a/2$ flips the Zak phase for each band: $\theta\rightarrow \theta+\pi$}.

\begin{figure}
\begin{centering}
\includegraphics[width=1\columnwidth]{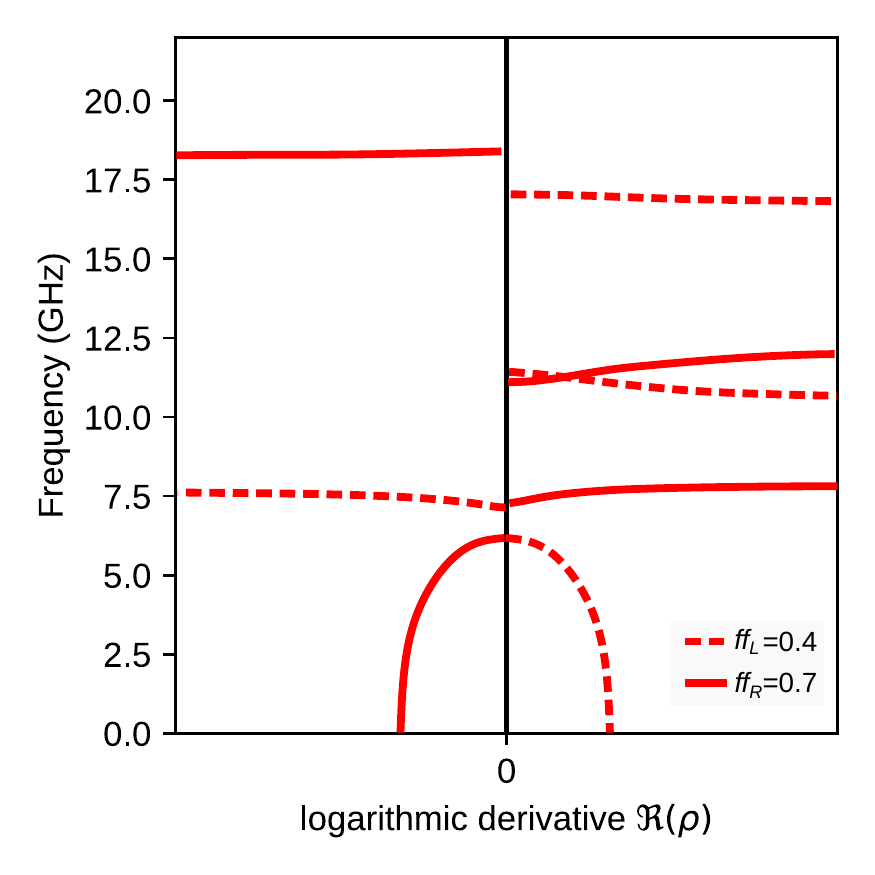}\caption{Logarithmic derivative calculated numerically from eq. (\ref{eq:log_derivative}). The lines (dashed for $f\!f=0.4$, solid for $f\!f=0.7$) are monotonically goes between 0 and infinity within the band gap. The crossing point for left and right MCs point out the frequency value of the interface state. Here, the interface state appears only in the second band gap.
}\label{fig:log_drv_04_07}
\par\end{centering}
\end{figure}

For the crystal with centrosymmetric unit cells, the properties of the Bloch function on the edges of bands are decisive for the sign of the logarithmic derivative of Bloch functions inside the gaps and thus determine the conditions of existence for surface/interface modes. According to the work of J. Zak \cite{Zak89}: (i) the logarithmic derivative  $\rho(k)$, taken in symmetry points $x_s=na$ or $x_s=a/2+na$, is real and has a constant sign in the whole range of the frequency gap, whereas in the band  $\rho(k)$ is purely imaginary; (ii) the sign of $\rho(k)$ in two successive gaps is different (the same) if $\rho(k)$ reaches two zeros or two poles (one zero and one pole) at the edges of the band between the gaps. The zeros and poles of $\rho(k)$ at the edges of bands correspond to $\left.\boldsymbol{m}_k\right|_{x=x_0}=0$ and  $\left.\partial_x\boldsymbol{m}_k\right|_{x=x_0}=0$, which means the odd and even Bloch functions at symmetry point corresponding to the edge of unit cell, respectively; (iii) the sign of $\rho$ is conserved (negated) in the direct gaps opened at $k_r=0$ ($k_r=\pi/a$), in the center of the BZ; (iv) the change of the side of interface $x_{0^+}\rightarrow x_{0^-}$ and the direction of decaying of the mode  from $x\rightarrow \infty$ to $x\rightarrow -\infty$ (when we switch from the MC on the right to the MC on the left) requires the change of the sign of imaginary part of the wavevector from $-k_i$ to $k_i$, that results in the change of the sign of the logarithmic derivative: $\left.\rho(k_r+i k_i)\right|_{x=x_{0^{+}}}=-\left.\rho(k_r-i k_i)\right|_{x=x_{0^{-}}}$. The properties (i), (ii) and conclusions highlighted at the end of the previous paragraph allow us to formulate the following statement. {\em When the Zak phase for a given band is equal $0$ ($\pi$) then the signs of logarithmic derivatives of Bloch function in gaps surrounding this band are  the same (are opposite).}

\begin{figure}
\begin{centering}
\includegraphics[width=1\columnwidth]{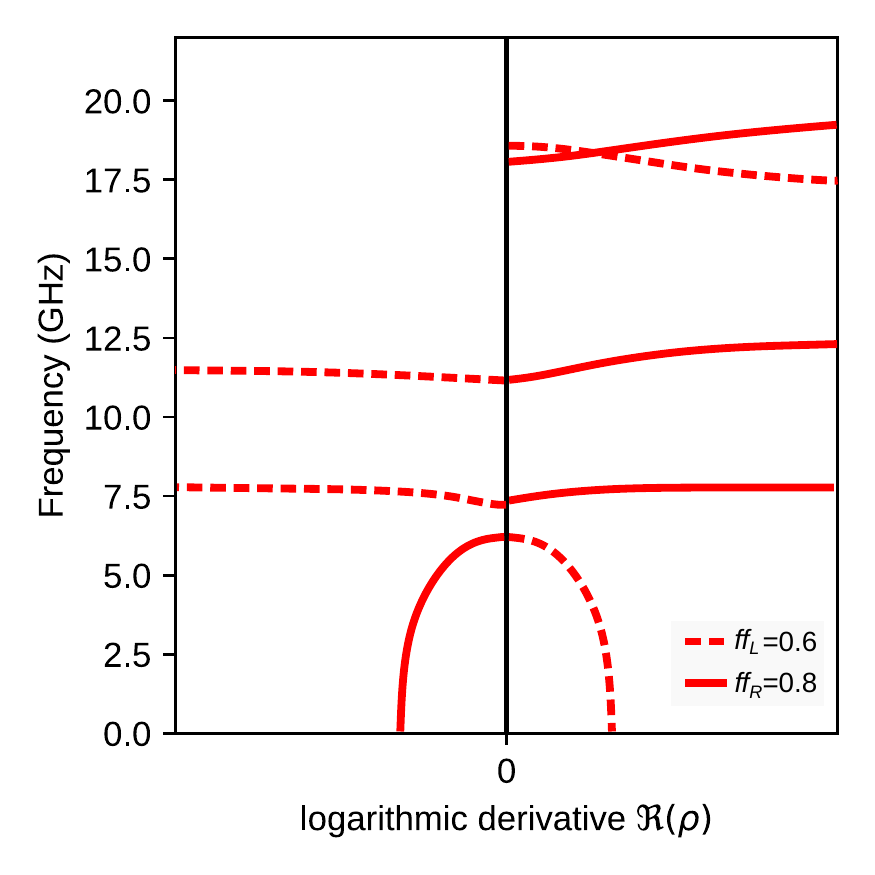}\caption{Logarithmic derivative calculated numerically from eq. (\ref{eq:log_derivative}). The lines (dashed for $f\!f=0.6$, solid for $f\!f=0.8$) are monotonically goes between 0 and infinity within the band gap. The crossing point for left and right MCs point out the frequency value of the interface state. Here, the interface state appears only in the third band gap.}\label{fig:log_drv_06_08}
\par\end{centering}
\end{figure}

The knowledge of the Zak phases for all bands below a given gap allows determining the sign of the logarithmic derivative of Bloch function $\rho(k)$ in this gap. To justify this formula, let us start with the discussion concerning the sign of $\rho(k)$ below the lowest band $n=1$.  For the frequencies below the $1^{\rm st}$ band, the Bloch function has a form of evanescent wave ($|k_i|>0$) with homogeneous phase ($k_r=0$) and without the oscillations within the unit cells. Therefore, it decays monotonously to the left $x\rightarrow -\infty$ ($k_i>0$) or to the right $x\rightarrow \infty$ ($k_i>0$) and its logarithmic derivative is positive or negative, respectively. This property is included by the sign $\pm$ in the formula which links the sign of $\rho(k)$ in the gap with the sequence of Zak phases for the bands below it (Eq.(5) in the manuscript). When all bands are characterized by the Zak phase $\theta=0$,  the logarithmic derivative of Bloch function $\rho$ flips its sign from the gap to the gap. Therefore, in the gap just above $n^{\rm th}$ band, it will be equal to $\pm(-1)^m$. The number of flips of the sign will be reduced if some bands with the Zak phase $\theta=\pi$ appear below the selected gap. If the odd (even) number of such bands exit, then the sign $\pm(-1)^m$  will (will not) be reversed. This potential reversal of the sign can be described by adding factor $\prod_{m=1}^n {\exp}(i\theta_m)= {\exp}\sum_{m=1}^n(i\theta_m)$, where the expression ${\exp}(i\theta_m)$ is equal to $+1$ or $-1$, if $\theta_m=0$ or $\pi$, respectively.

\section{Calculations of the logarithmic derivative of Bloch function using the PWM}
\label{sec:App.C}

Using the Fourier expansion of the Bloch function (\ref{eq:FT}), we can strictly calculate the derivative $\partial_x   m_{\parallel(\perp)}$ and write the logarithmic derivative of Bloch function in the form:

\begin{equation}
    \rho(x)=i\frac{\sum_i (k+G_i)u_{\parallel(\perp),k,G_i}\exp({iG_i x})}{\sum_n u_{\parallel(\perp),k,G_i}\exp({iG_i x})}.\label{eq:log_derivative}
\end{equation}

\section{The exemplary SWs' profiles on the edges of the frequency bands }
\label{sec:App.D}

 For the centrosymmetric unit cell, the Zak phase $\theta_n$ of each $n^{\rm th}$ band of the 1D crystal can be determined only by inspection of the Bloch functions' symmetry  $m_{k,\parallel(\perp),n}\left(x\right)$ in the symmetry point of the $1^{\rm st}$ Brillouin zone: $k=0$ or $k=\pm \pi/a$, where $a$ is a period of 1D lattice. The Zak phase of the $n^{th}$ band is $0$ if either
$\left|m_{n,k=0}\left(x=x_0\right)\right|=\left|m_{n,k=\pi/a}\left(x=x_0\right)\right|=0$ or $\left|\partial_{x} m_{n,k=0}\left(x=x_0\right)\right|=\left|\partial_{x} m_{n,k=\pi/a}\left(x=x_0\right)\right|=0$, where $x_0$ denotes the center of the unit cell. Otherwise, it is $\pi$\cite{Xiao14}.

Figs. \ref{fig:vec_full}(a, b) presents the profiles of modes for four the lowest bands for the pair of $f\!f=0.4$ and 0.7 (up and down, respectively) and  Fig.~\ref{fig:vec_full}(c, d) --for 0.6, 0.8, and Fig.~\ref{fig:vec_full}(e, f) for the system with included dipolar interaction: $f\!f$=0.6 and 0.9.
Black lines marks the profiles for $k=0$, and green lines -- for $k=\pi/a$. Light blue and yellow colors denotes the regions in which the Py and Co strips are placed, respectively (see Fig.~1 in the manuscript). The Zak phase can be deduced by analyzing symmetry of the profiles and the edges of each band, i.e., for $k=0$ and $k=\pi/a$.  

\begin{widetext}

\begin{figure}[ht]
\centering
\includegraphics{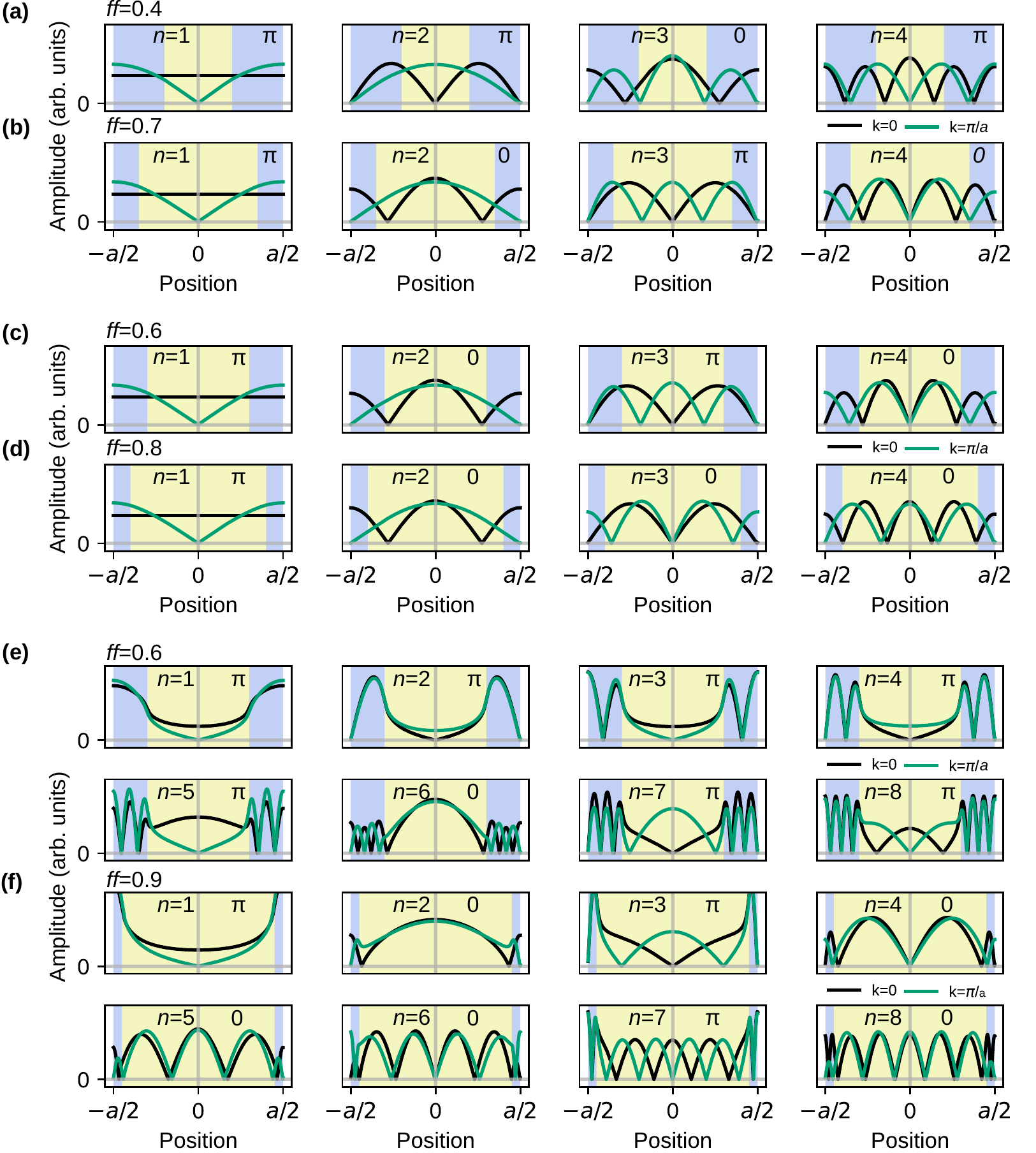}
\caption{SW's profiles $|m_{\bot}|$ within the unit cell at the edges of BZ. Black line is calculated for $k=0$, green line for $k=\pi/a$. The values of the SW profiles and their derivatives allow determining the Zak phase for the band. 
(a) Results for $f\!f=0.4$. For $n=1$, $n=2$ and $n=4$ modes have different derivatives in the middle of unit cells, that suggest value of Zak phase equal to $\pi$. For $n=3$ Zak phase is equal to 0. 
(b) Results for $f\!f=0.7$. For $n=1$ and $n=3$ Zak phase is equal $\pi$, and for $n=2$ and $n=4$ Zak phase is equal 0. 
(c) Results for $f\!f=0.6$. For $n=1$ and $n=3$ Zak phase equal to $\pi$. For $n=2$ and $n=4$ Zak phase is equal to 0. 
(d) Results for $f\!f=0.8$. Only for $n=1$ Zak phase is equal $\pi$, everywhere else is equal to 0.
For exchange-dipolar waves, 8 SW's profiles are presented. 
(e) Results for $f\!f=0.6$. For $n=1$, $n=2$, $n=3$, $n=4$, $n=5$, $n=8$ Zak phase is equal to $\pi$, everywhere else is equal to 0. 
(f) Results for $f\!f=0.9$. For $n=1$, $n=3$, $n=7$ Zak phase is equal to $\pi$, everywhere else is equal to 0. 
\label{fig:vec_full}}
\end{figure}   

\clearpage
\end{widetext}

 \bibliography{bibliography}

\end{document}